\documentclass[journal,twocolumn]{IEEEtran}

\usepackage{xcolor}
\usepackage[normalem]{ulem}

\newcommand{\sps}{SpS}
\newcommand{\sab}{SAB}
\newcommand{\bilstm}{BiLSTM}

\newcommand{\bilstmfc}{BiLSTM-FC}
\newcommand{\convfc}{Conv-FC}

\newcommand{\vect}[1]{\boldsymbol{#1}}
\newcommand{\qat}{QAT}

\newcommand{\myxor}{\oplus}
\newcommand{\myor}{\vee}
\newcommand{\myand}{\wedge}

\newcommand{\eg}{e.g.,} % not italic 
\newcommand{\ie}{i.e.,}

\usepackage{algorithm}
\usepackage{algpseudocode}
\usepackage{amsmath,amsfonts,amssymb}
\usepackage{bbm}

\usepackage{xcolor}
\definecolor{darkblue}{RGB}{0,0,128}
\definecolor{darkred}{RGB}{128,0,0}
\definecolor{darkgreen}{RGB}{0,128,0}
\definecolor{darkbrown}{RGB}{153,76,0}
\definecolor{boxblue}{RGB}{175,238,238}

\definecolor{dark}{RGB}{175,238,238}

\usepackage{cite}

\usepackage{tikz,pgfplots}
\usepackage{pgfplotstable}

\DeclareMathOperator{\sign}{sign}
\DeclareMathOperator{\one}{\mathbbm{1}}
\DeclareMathOperator{\clip}{\textnormal{clip}}
\DeclareMathOperator{\erfc}{erfc}

\DeclareUnicodeCharacter{2212}{-}

\usepackage{hhline}

\usepackage{caption} 
\usepackage{flushend}

\begin{document}

\title{Quantization of Neural Network Equalizers in Optical Fiber Transmission Experiments}

\author{
Jamal Darweesh, Nelson Costa, Antonio Napoli, Bernhard Spinnler, Yves Jaouen, and Mansoor Yousefi
\thanks{
Manuscript submitted September, 2023. 
This work has received funding from the European Union's Horizon 2020 research and innovation programme 
under the Marie Sk\l odowska-Curie Grant Agreement no. 813144,  and the European Research Council (ERC) research and
innovation programme, under the COMNFT project, Grant Agreement no. 805195.

Jamal Darweesh, Yves Jaouen and Mansoor Yousefi are with Telecom Paris, Institut Polytechnique de Paris, 91120 Palaiseau,
France (e-mail:  \{jamal.darweesh, yves.jaouen, yousefi\}@telecom-paris.fr). 

Nelson Costa is with Infinera Unipessoal, 2790-078 Carnaxide, Portugal (e-mail: ncosta@infinera.com).

Bernhard Spinnler and Antonio Napoli are with Infinera, 81541 Munich, Germany (e-mail: \{anapoli, bspinnler\}@infinera.com).

Copyright~\copyright~2018 IEEE. Personal use of this material is permitted. However, permission to use this 
material for any other purposes must be obtained from the IEEE by sending a request to pubs-permissions@ieee.org.
}
}

%\markboth{September 2023}%
%{Quantization of Neural Networks for Nonlinearity Mitigation in Fiber Transmission Experiment}

%\IEEEoverridecommandlockouts
%\IEEEpubid{\makebox[\columnwidth]{978-1-5386-5541-2/18/\$31.00~\copyright2018 IEEE \hfill} \hspace{\columnsep}\makebox[\columnwidth]{ }}

\maketitle

%\IEEEpubidadjcol

\begin{abstract}
The quantization of neural networks for the mitigation of the nonlinear and components' distortions in  dual-polarization 
optical fiber transmission is studied. 
Two low-complexity neural network equalizers
are applied in three 16-QAM 34.4 GBaud transmission experiments with  different representative fibers.
A number of post-training quantization and quantization-aware training algorithms are compared for casting 
the weights and activations of the neural network in few bits, combined with the uniform, additive power-of-two, 
and companding quantization. 
For quantization in the large bit-width regime of  $\geq 5$ bits, the quantization-aware training  with the 
straight-through estimation incurs a Q-factor penalty of less than 0.5 dB compared to the unquantized neural network.
For quantization in the low bit-width regime, an algorithm dubbed companding successive alpha-blending  quantization is suggested.
This method compensates for the quantization error aggressively by successive grouping and retraining of the parameters, 
as well as an incremental transition from the floating-point representations to the quantized values within each group.
The activations can be quantized at 8 bits and the weights on average at 1.75 bits,  with a penalty of $\leq 0.5$~dB. 
If the activations are quantized at 6  bits, the weights can be quantized at 3.75 bits with minimal penalty.  
The computational complexity and required storage of the neural networks are drastically reduced, typically by over 90\%.
The  results indicate that low-complexity neural networks can mitigate nonlinearities in optical fiber transmission.

\end{abstract}


%%%%%%%%%%%%%%%%%%%%%%%%%%%%%%%%%%%%
%%%%% SECTION I: Introduction
%%%%%%%%%%%%%%%%%%%%%%%%%%%%%%%%%%%

\def\sfactor{0.85}
\section{Introduction}

\IEEEPARstart{T}{he} compensation of the channel impairments is essential to the spectrally-efficient optical fiber transmission.  The
advent of the coherent receivers, combined with the advances in the digital signal processing (DSP) algorithms, has
allowed for the mitigation of the fiber transmission effects in the electrical domain \cite{savory2010digital}.  However,
real-time energy-efficient DSP is challenging in high-speed communication.

The linear transmission effects, such as the chromatic dispersion (CD) and polarization mode dispersion (PMD), can be
compensated using the well-established DSP algorithms \cite{agrell2016roadmap}.
The distortions arising from the fiber Kerr nonlinearity can in principle be partially compensated using  the digital
back propagation (DBP) based on the split-step Fourier method (SSFM).
DBP can be computationally complex in long-haul transmission with large number of steps in
distance \cite{dar2017}. 
The neural networks (NNs) provide an alternative approach to nonlinearity mitigation with flexible
performance-complexity trade-off  \cite{gibson1989,ibnkahla2000,jarajreh2015,zhang2018,butler2021}; see Section \ref{sec:lit}.

To implement NNs for real-time  equalization, the model should be carefully optimized for the
hardware.
The number of bits required to represent the NN can be minimized by quantization \cite{vanhoucke2011} and 
data compression, using techniques such as pruning, weight sharing and clustering \cite{han2016compress}.
There is a significant literature showing that these methods often drastically reduce the storage requirement of the NN, 
and its energy consumption, which is often dominated by the communication cost of
fetching words from the memory to the arithmetic units \cite{han2016compress,hubara2017quant,jacob2018quant}.
How the NNs can be quantized with as few bits as possible,  while maintaining a given Q-factor, is an important problem. This
paper is dedicated to the quantization of the NNs for nonlinearity mitigation, in order to reduce the computational 
complexity, memory footprint, latency and energy consumption of the DSP.

There are generally two approaches to the NN quantization. 
In post-training quantization (PTQ), the model is trained in 32- or 16-bit floating-point (FP) precision, and the resulting 
parameters are then quantized with fewer number of bits \cite{vanhoucke2011,banner2019post}. This  approach is  simple; however, 
quantization introduces a perturbation to the model parameters incurring a performance penalty. 
As a consequence, PTQ is usually applied in applications that do not require quantization below 8 bits.

In quantization-aware training (\qat), quantization is integrated into the training algorithm, and the quantization
error is partly compensated \cite{hubara2017quant,courbariaux2015bc,courbariaux2016binarized,jacob2018quant,krishnamoorthi2018}.
However the optimization of the loss function with gradient-based methods is not directly possible, because 
the quantizer has a derivative that is zero almost everywhere.
In the straight-through estimator (STE), the quantizer
is assumed to be the identity function, potentially saturated  in an input interval, in 
the backpropagation algorithm used for computing the gradient of the loss function \cite{hinton2012course,bengio2013ste}.
\qat\ is used in applications requiring low complexity in inference; 
however, it can be more complex in training than PTQ, and needs parameter tuning and experimentation. 
With the exception of a few papers reviewed in Section~\ref{sec:q-review},
the quantization of the NNs for nonlinearity mitigation has not been much explored.

\begin{figure*}
%\vspace{-11pt}
\centering
\includegraphics[width=0.99\textwidth]{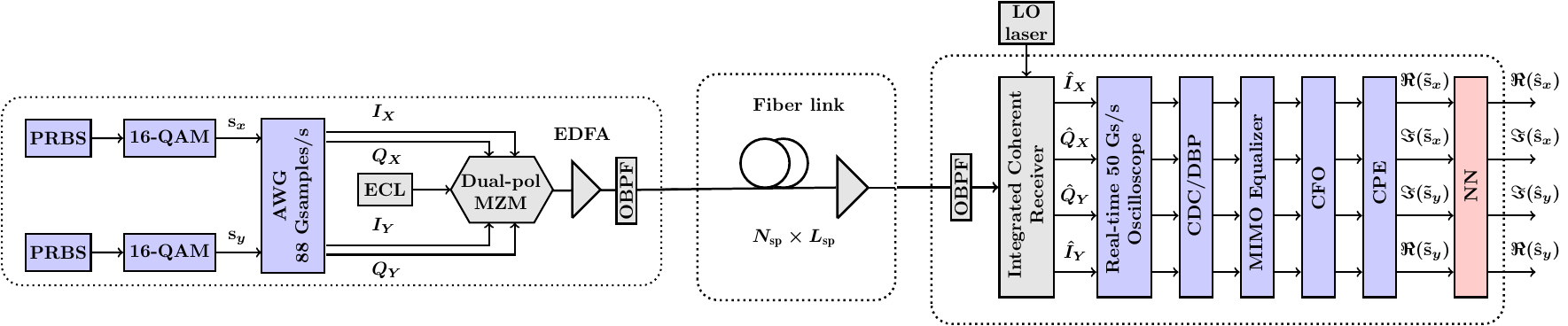}
\caption{The block-diagram of the transmission experiments.} 
\label{fig:sys-model}
\end{figure*}

In this paper, we study the quantization of the weights and activations of 
a small convolutional fully-connected (\convfc) and a bidirectional long short-term memory fully-connected (\bilstmfc) equalizer,
applied to three  16-QAM 34.4 GBaud dual-polarization fiber transmission experiments.
The experiments are based on a 9x50 km true-wave classic (TWC) fiber link, a 9x110 km standard single-mode fiber (SMF) link, and a
17x70 km large effective area fiber (LEAF) link.
We compare the Q-factor penalty, computational complexity, and memory requirement of a number of PTQ and \qat-STE algorithms,
as a function of the launch power and the quantization rate $b$.
The uniform, additive power-of-two (APoT), companding,   fixed- and mixed-precision 
quantization are compared. It is shown that, these algorithms, if optimized, work well in the large bit-width regime of $b\geq 5$.
However,  they do not achieve sufficiently small distortions in our experiments in the low bit-width regime with $b<5$, 
where the quantization error needs to be aggressively mitigated.
For this case,
we propose a companding successive  alpha-blending  (\sab) quantization algorithm that 
mitigates the quantization error by successive grouping and retraining of
the parameters, combined with an incremental transition from the floating-point representations to the quantized values within each
group. 
The algorithm also accounts for the probability distribution of the parameters.
It is shown that the quantization of the activations impacts the Q-factor much more than the weights. 
The companding \sab\ algorithm is studied w/wo the quantization of activations.

The results indicate that, for quantization in the large bit-with regime, \qat-STE incurs 
a Q-factor penalty of less than 0.5 dB relative to the unquantized NN, while reducing the
storage and computational complexity of the NN typically by over
90\%. This is obtained with the uniform, companding or APoT variant of \qat-STE, depending on the transmission experiment.
If the activations are quantized at 8 bits, the weights can be quantized with the companding \sab\ algorithm at the average rate
of 1.75 bits, paving the way to the binary NN equalizers.
The quantization of the activations at 6 bits and weights at $3.75$ bits results in 
a reduction in the computational complexity by $95\%$ and memory  footprint by $88\%$, with the Q-factor penalty of 0.2 dB.
Overall, the results suggest that nearly-binary NNs  mitigate nonlinearities in optical fiber transmission.

This paper is structured as follows. In Section \ref{sec:dual-pol-model}, 
we describe the optical fiber transmission experiments. 
In Section \ref{sec:nns-NL}, we review the use of the NNs for the fiber nonlinearity mitigation, and in Section
\ref{sec:quant} the quantization of the NNs. 
Finally, we compare the Q-factor penalty and the gains of quantization for several algorithms 
in Section \ref{sec:results}, and draw conclusions in Section~\ref{sec:conc}.

%%%%%%%%%%%%%%%%%%%%%%%%%%%%%%%%%%%%
%%%%% SECTION II: Dual Polarization Transmission experiment setup
%%%%%%%%%%%%%%%%%%%%%%%%%%%%%%%%%%%

\section{Dual Polarization Transmission Experiment Setup}
\label{sec:dual-pol-model}

Fig.~\ref{fig:sys-model} shows the block diagram of the transmission experiments 
considered in this paper. 
Three experiments are performed with different representative fibers, described below.

\subsubsection{Transmitter}

At the transmitter (TX), a pseudo-random bit sequence (PRBS) is generated for each polarization  $p\in\{x,y\}$, and
mapped to a sequence of symbols $\mathbf s_p$  taking values in a 16-QAM constellation according to the Gray mapping. 
The two complex-valued sequences $\mathbf s_x$ and $\mathbf s_y$ are converted to four real-valued sequences,
and passed to an arbitrary wave generator (AWG)
that modulates them to two QAM signals using a root raised cosine  pulse shape with the roll-off factor of 0.1
at the rate $34.4$ GBaud. 
The AWG includes digital-to-analog converters (DACs) at $88$  Gsamples/s.

The outputs of AWG are four continuous-time electrical signals  $I_x$, $Q_x$, $I_y$ and $Q_y$ corresponding 
to the in-phase (I) and quadrature (Q) components of the signals of the $x$ and $y$ polarization.
The electrical signals are converted to optical signals and polarization-multiplexed with a dual-pol IQ Mach-Zehnder 
modulator (MZM), driven by an external cavity  laser (ECL) at wavelength  $1.55~\mu m$ with line-width 100 KHz.
The output of the IQ-modulator is amplified by an erbium-doped fiber amplifier (EDFA), filtered by an optical
band-pass filter (OBPF) and launched into the fiber link.
The laser introduces phase noise, modeled by a Wiener process with the Lorentzian power spectral density 
\cite[Chap.~3.5]{agrawal2021}.

\def\lsep{1pt}
\setlength{\arrayrulewidth}{0.8pt}
\begin{table}[t]
    \caption{\MakeUppercase{Optical link parameters}\label{tab:links}}
\centering
\begin{tabular}{l c@{~~~~~~}c c }
                           & TWC  fiber &  SMF & LEAF  \\[\lsep]
                           \cline{2-2} \cline{3-3} \cline{4-4}\\[-6pt] 
$L_{\textnormal{sp}}$ ${\rm km}$                  & 50          & 110        & 70 \\[\lsep]
$N_{\textnormal{sp}}$                   & 9          & 9        & 17 \\[\lsep]
$\alpha$ ${\rm dB/km}$                      & 0.21          & 0.22        & 0.19 \\[\lsep]
$D$ ${\rm ps/(nm.km)}$                      & 5.5          & 18        & 4 \\[\lsep]
$\gamma$ ${\rm (W.Km)^{-1}}$                  & 2.8          & 1.4        & 2.1 \\[\lsep]
PMD $\tau$ $\rm ps/\sqrt{km}$                      & 0.02         & 0.08      &0.04 \\[\lsep]
NF     ${\rm dB}$              & 5          &     5    & 5\\
%$G$     ${\rm dB}$              &           &         &  \\
\end{tabular}
\end{table}

\subsubsection{Fiber-optic Link}

The channel is a straight-line optical fiber link in a lab, with $N_{sp}$ spans of length $L_{sp}$. An EDFA with 
5 dB noise figure (NF) is placed at the end of each span to compensate for the fiber loss. 
The experiments are performed  with the TWC fiber, SMF and LEAF, and parameters in Table \ref{tab:links}.

\paragraph{TWC Fiber Experiment}
\label{sec:TWC-exp}

The first experiment is  with a short-haul TWC
fiber link with 9 spans of 50 km. The TWC fiber was a brand of nonzero dispersion shifted fiber (NZ-DSF)  made by Lucent, with low 
CD coefficient of $D=5.5~{\rm ps/(nm\cdot km)}$ at 1550 nm wavelength and a high nonlinearity parameter
of $\gamma=2.8~({\rm Watt\cdot km})^{-1}$. Thus, even though the link is short with 450 km length, 
the channel operates in the nonlinear regime at high powers. The link parameters, including the fiber
loss coefficient $\alpha$ and PMD value $\tau$,  can be found in Table \ref{tab:links}.

\paragraph{SMF Experiment}
\label{sec:SMF-exp}

The second experiment is based on a long-haul 9x110 km standard single-mode fiber  link, 
with parameters in Table \ref{tab:links}.

\paragraph{LEAF Experiment}
\label{sec:LEAF-exp}

LEAF is also a brand of NZ-DSF, made by Corning, similar to the TWC fiber but with a smaller nonlinearity
coefficient due to the larger cross-section effective area. This experiment uses a 17x70 km link described in Table \ref{tab:links}.

\subsubsection{Receiver}
\label{sec:rx}

At the receiver,  the optical signal is polarization demultiplexed, and 
converted to four electrical signals  using an integrated coherent receiver driven by a local oscillator (LO).
Next, the continuous-time electrical signals are converted to the discrete-time signals by an oscilloscope, 
which includes analog-to-digital converters (ADCs) that sample the signals at the rate of $50$  Gsamples/s, 
and quantize them with the effective number of  bits of around $5$.
The digital signals are up-sampled at 2 samples/symbol, and equalized in the DSP chain shown in
Fig.~\ref{fig:sys-model}.

The equalization is  performed  by the conventional dual-polarization linear DSP \cite{savory2010digital}, 
followed by a NN. The linear DSP consists of a cascade of the frequency-domain CD compensation, multiple-input 
multiple-output (MIMO) equalization via the radius directed equalizer  
to compensate for PMD  \cite[Sec. VII-]{savory2010digital}, \cite{fatadin2009}, 
polarization separation, carrier frequency offset (CFO) correction,
and the carrier-phase estimation (CPE) using the two-stage   algorithm
of Pfau \emph{et al.} to compensate for the phase offset \cite{pfau2009PN}. 
The linearly-equalized symbols are denoted by $\tilde{\mathbf{s}}_p$.

Once the linear DSP is applied, the symbols  are still subject to the
residual CD, dual-polarization nonlinearities, and the distortions introduced by the components at TX and RX.
Define the residual channel memory $M$ to be the maximum effective length of the auto-correlation function of 
$\tilde{\mathbf{s}}_p$ over $p\in\{x,y\}$.

The outputs of the CPE block $\tilde{\mathbf{s}}_p$  are passed to a low-complexity NN, which mitigates 
the remaining distortions, and outputs $\hat{\mathbf{s}}_p$.
The architecture of the NN depends on the experiment, and will be 
explained in Section~\ref{sec:two-nns}.

%%%%%%%%%%%%%%%%%%%%%%%%%%%%%%%%%%%%
%%%%% SECTION III: Neural networks for nonlinearity mitigation
%%%%%%%%%%%%%%%%%%%%%%%%%%%%%%%%%%%

\section{Neural Networks for Nonlinearity Mitigation}
\label{sec:nns-NL}

\subsection{Prior Work}
\label{sec:lit}

The NN equalizers in optical fiber communication can be classified into two categories. 
In \emph{model-based equalizers}, the architecture is based on the parameterization of the channel model.
An example is learned DBP (LDBP) \cite{butler2021}, where the NN is a parameterization of the SSFM which is often 
used to simulate the fiber channel. 
The dual-polarization LDBP is a cascade of layers, each consisting of two complex-valued symmetric filters to compensate for the CD, two real-valued
asymmetric filters for the differential group delays, a unitary matrix for the polarization rotation, and a 
Kerr activation function for the mitigation of the fiber nonlinearity. 
It is shown that LDBP outperforms DBP  \cite{butler2021}.

On the other hand, in \emph{model-agnostic equalizers}, the architecture is independent of the channel
model \cite{gibson1989,ibnkahla2000,jarajreh2015,zhang2018}.
The model-agnostic schemes do not require the channel state information, such as the fiber parameters.
Here, the  NNs can be placed at the end of the conventional linear DSP for nonlinearity mitigation
\cite{sidelnikov2018equalization}, or  after the ADCs for compensating 
the linear and nonlinear distortions (thereby replacing the linear DSP) \cite{Karanov:18,shahkarami2022complexity}.

A number of NN architectures have been proposed for the nonlinearity mitigation. 
Fully-connected (FC) or dense NNs  with 2 or 3  layers, 
few hundred neurons per layer, and tanh activation were studied in
\cite{Freire:21,catanese2020fully}.
The overfitting and complexity become problems when the models get bigger. 
The convolutional NNs can model the linear time-invariant (LTI) systems with a finite impulse response.
The application of the convolutional networks for compensating the nonlinear distortions is investigated in
\cite{Sidelnikov:21}, showing that one-dimensional convolution can well  compensate the CD.
The bi-directional recurrent and long-short term memory networks (LSTM) receivers 
are shown to perform well in fiber-optic equalization \cite{shahkarami2022complexity}. 
Compared to the convolutional and dense networks,  \bilstm\ networks better model LTI systems with infinite impulse 
response, such as the response of the CD.
A comparison of the different architectures in optical transmission in \cite{Freire:21} shows that,  
dense and convolutional-LSTM models perform well at low and high complexities, respectively.

\begin{figure*}[t]
\begin{center}
\begin{tabular}{c@{~~~~}c}
%\begin{tabular}{ccc}
\includegraphics[width=0.65\textwidth]{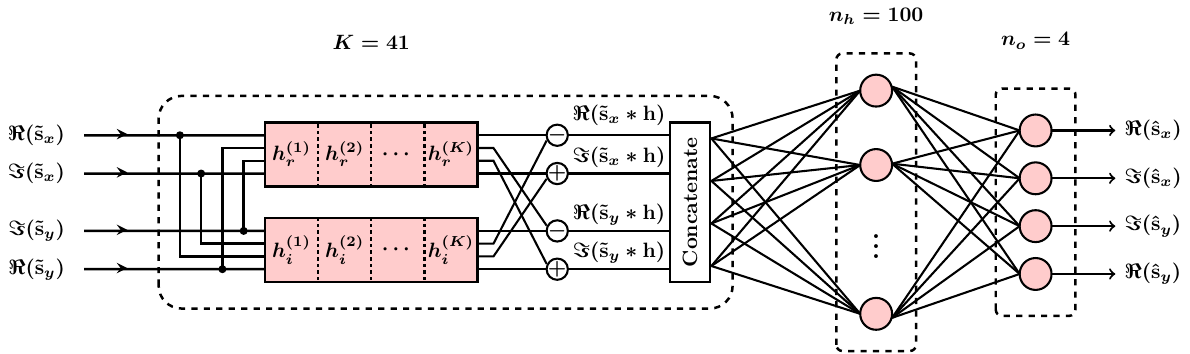}
& 
\includegraphics[width=0.335\textwidth]{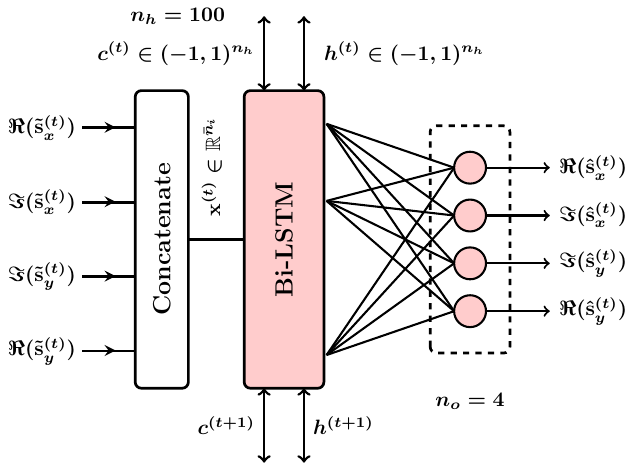}\\
 {\fontsize{8}{12}\selectfont  (a)} & {\fontsize{8}{12}\selectfont (b) } 
\end{tabular}
\end{center}
\caption{Architectures of the NN. The input is the linearly-equalized symbols $\tilde s_x$ and $\tilde s_y$, and the 
output is the fully-equalized symbols $\hat s_x$ and $\hat s_y$. (a) \convfc\ model. The convolutional filter taps are
indicated by $h_r^{(l)} =\bigl[\Re(\mathbf h)\bigr]_{l}$ and $h_i^{(l)} =\bigl[\Im(\mathbf h)\bigr]_{l}$; (b) \bilstmfc\ model.}
\label{fig:arcs}
\end{figure*}

An effect that particularly impacts the performance of the NN is PMD. In most papers, 
random variation of the polarization-dependent effects during the transmission have not been carefully studied. 
The polarization effects are sometimes neglected \cite{sidelnikov2018equalization}, or assumed to 
be static during the transmission \cite{butler2021}. In such simulated systems, the dual-polarization 
NN receivers are subject to a performance degradation compared to real-life experiments \cite{Freire:21}.

\subsection{Two NN Models Considered in This Paper}
\label{sec:two-nns}

In this Section, we describe two NN equalizers used in this paper. 
The NN is placed at the end of 
the linear DSP shown in Fig.~\ref{fig:sys-model}. In consequence, since the PMD is compensated by the 
MIMO equalizer, the NN is static and trained offline. 
Due to the constrains of the practical systems, low-complexity architectures are considered. 
A \convfc\ network is applied in the  TWC fiber and SMF links, and a \bilstmfc\ network in the LEAF link.
The \bilstmfc\ model has more parameters, and performs better; however,  the smaller \convfc\ model is sufficient
in short-haul links.

\subsubsection{\convfc\ Model}
\label{sec:conv-fc}

The four sequences of linearly-equalized symbols $\Re(\tilde{\mathbf{s}}_x)$, $\Im(\tilde{\mathbf{s}}_x)$, $\Re(\tilde{\mathbf{s}}_y)$ 
and $\Im(\tilde{\mathbf{s}}_y)$ are passed to the NN.
We consider a many-to-one architecture, where the NN equalizes one complex symbol per polarization given $n_i$ input symbols. 
The inputs of the network are four vectors, each containing a window of $n_i=M+1$ consecutive elements from each of the four
input sequences, 
where $M$ is the residual channel memory defined in Section \ref{sec:rx}. The network outputs a vector of  $n_o=4$ real numbers,
corresponding to the real and imaginary parts of the symbols of the two polarizations after full
equalization.
The  size of the concatenated input of the NN  is  thus $\bar{n}_i=4(M+1)$.
The NN operates in a sliding-window fashion: as each of its input vectors are shifted forward 
one element,  $4$ real numbers are produced.

\def\sfactor{0.3}
\begin{figure*}[t]
\begin{center}
\begin{tabular}{c@{~~~~~}c@{~~~~~}c}
\includegraphics[width=\sfactor\textwidth]{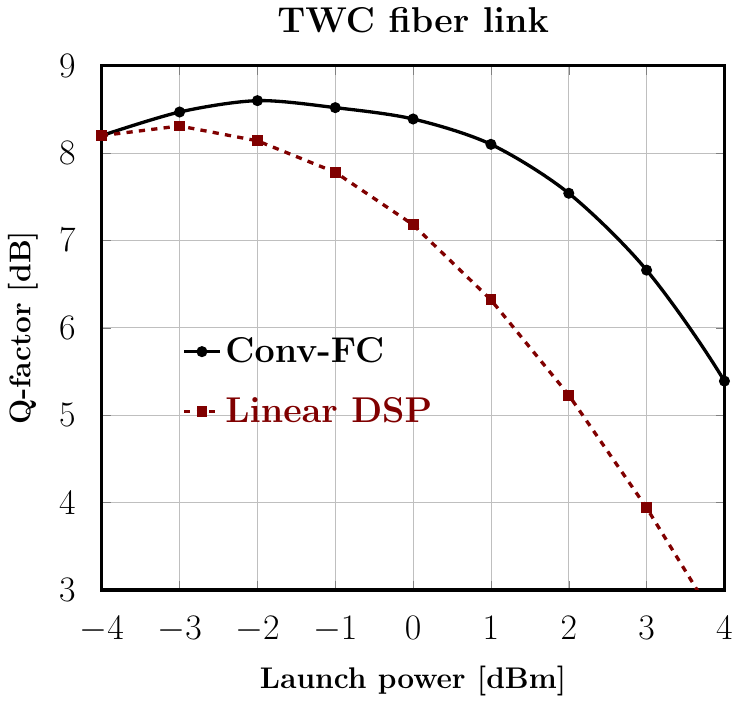}&
\includegraphics[width=\sfactor\textwidth]{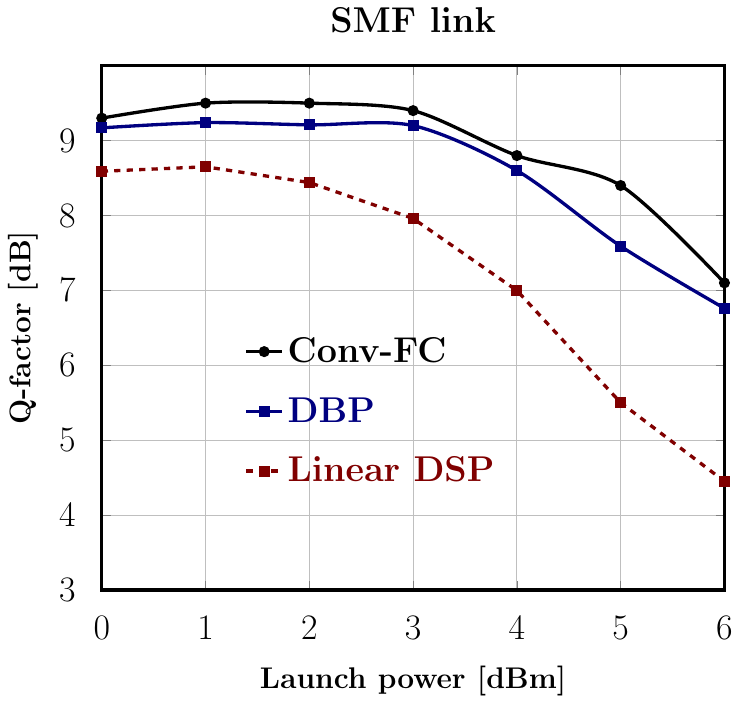}&
\includegraphics[width=\sfactor\textwidth]{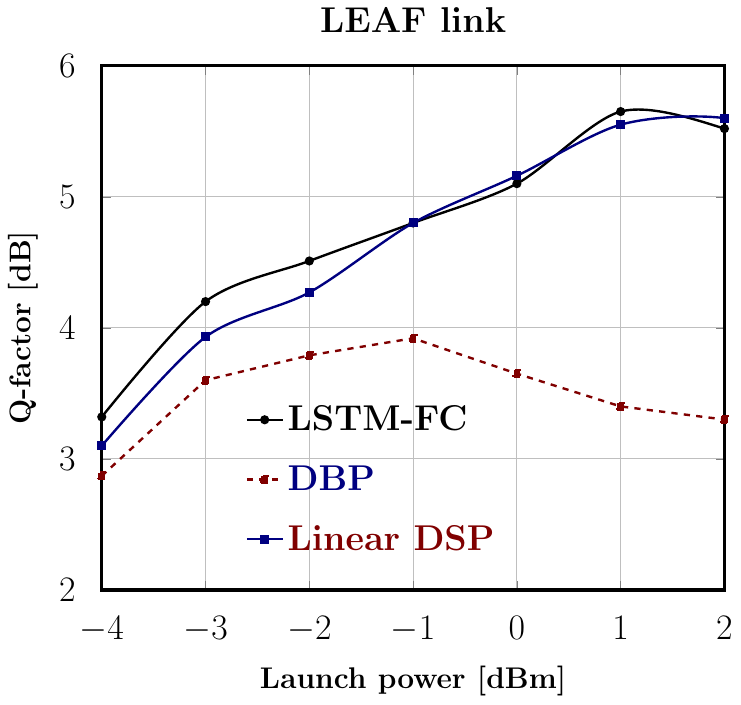}
\\
{\fontsize{8}{12}\selectfont (a)} & {\fontsize{8}{12}\selectfont (b)} &{\fontsize{8}{12}\selectfont  (c)} 
\end{tabular}
\end{center}
\caption{ Q-factor of the linear DSP, DBP with 1 \sps, and  unquantized NN equalizers  in (a)  TWC fiber, (b) SMF, and (c) LEAF
experiments.}
\label{fig:perf-uq}
\end{figure*}

The \convfc\ model is a cascade of a complex-valued convolutional layer, a FC hidden layer, and a FC output layer. 
The first layer implements the discrete convolution 
of $\tilde{\mathbf{s}}_p$, $p\in\{x,y\}$, with a kernel $\mathbf h \in\mathbb{C}^K$, to compensate primarily 
the residual CD, where $\mathbb C$ denotes the complex numbers and $K$ is the number of kernel 
taps.
The two  complex convolutions  $\tilde{\mathbf{s}}_p * \mathbf{h}$ are implemented using eight real convolutions in
terms of two filters
$\Re(\mathbf h)$ and $\Im(\mathbf h)$, according to
\begin{IEEEeqnarray}{rCl}
    \tilde{\mathbf{s}}_p * \mathbf{h} &=&  \Re(\tilde{\mathbf{s}}_p) * \Re(\mathbf{h})- \Im(\tilde{\mathbf{s}}_p) * \Im(\mathbf{h})
\nonumber\\
&&+j\Bigl\{ \Re(\tilde{\mathbf{s}}_p) * \Im(\mathbf{h})+\Im(\tilde{\mathbf{s}}_p) * \Re(\mathbf{h})\Bigr\}.
\label{eq:conv}
\end{IEEEeqnarray}
The first layer thus contains eight parallel real-valued one-dimensional convolutions,  with the stride one
and  ``same padding,'' and no activation.
There are total $2K$ trainable real filter taps, typically far fewer than in generic 
convolutional layers used in the literature with large feature maps. 
The eight real convolutions are combined according to (\ref{eq:conv}) or Fig.~\ref{fig:arcs}(a), obtaining 
$\Re(\tilde{\mathbf{s}}_x * \mathbf{h})$, $\Im(\tilde{\mathbf{s}}_x * \mathbf{h})$, 
$\Re(\tilde{\mathbf{s}}_y * \mathbf{h})$ and $\Im(\tilde{\mathbf{s}}_y * \mathbf{h})$, 
which are  then concatenated. The resulting vector is fed to a  FC hidden layer with $n_h$ neurons, and tangent 
hyperbolic (tanh) activation. The joint processing of the two polarizations in the dense layer is 
necessary in order to compensate  the nonlinear interactions between the two polarizations during the 
propagation. Finally, there is an output FC layer with $2$ neurons for each complex-valued polarization symbol, and
no activation.

The computational complexity $\mathcal C$ of the unquantized NNs can be measured by the number of the real multiplications 
per polarization,  considering that the cost of 
the additions and  computation of the activation is comparatively negligible. 
For the \convfc\ model 
\begin{IEEEeqnarray}{rCl}
\mathcal{C}_{\textnormal{\convfc}}= 4n_iK +2n_in_h+  \frac{n_hn_o}{2}.
\label{eq:complexity}
\end{IEEEeqnarray}

\subsubsection{\bilstmfc\ Model}
\label{sec:lstm-fc}

The  second model is a cascade of a  concatenator,  a \bilstm\ unit and FC output layer, shown in Fig.~\ref{fig:arcs}(b).
At each time step $t$ in the recurrent model, $n_i=M+1$ linearly-equalized complex symbols are taken from each polarization.
The resulting vectors $\Re(\tilde{\mathbf{s}}_x^{(t)})$, $\Im(\tilde{\mathbf{s}}_x^{(t)})$, $\Re(\tilde{\mathbf{s}}_y^{(t)})$, 
$\Im(\tilde{\mathbf{s}}_y^{(t)})$ 
are concatenated in a vector of length $\bar n_i=4(M+1)$  and fed to a many-to-many \bilstm\ unit.
Each LSTM cell in this unit has an input of length $2(M+1)$ corresponding to the one-sided memory, 
$n_h$ hidden state neurons, the  recurrent activation $\tanh$, and the gate activation sigmoid. 
The output of the \bilstm\ unit is a vector of length $2n_h$, that is fed  to a FC output layer with 
no activation and $n_o=4$ neurons\footnote{Equivalently, the input output of the \bilstm\ unit may be expressed in arrays of 
shape $(4,M+1)$, without concatenation.}.
The computational complexity of the \bilstmfc\ model is
\begin{equation*}
   C_{\textnormal{\bilstmfc}}= n_{h}\Bigl(4n_{h}+16n_i+3+n_o \Bigr),
\end{equation*}
real multiplications per polarization. 

The many-to-many variants of the above models are straightforward. In this case,
there are $n_o=4(M+1)$ neurons at the output, so that all $M+1$ complex symbols are equalized in one
shot; 
thus  $n_i=M+L$, $\bar{n}_i=n_o=4(M+L)$. The many-to-many versions are less complex per symbol and parallelizable, but also less performant.

The performance of the receiver is measured in terms of 
\begin{IEEEeqnarray*}{rCl}
\textnormal{Q-factor}=10\log_{10}\Bigl(2\erfc^{-2}(2\textnormal{BER})\Bigr)\quad \textnormal{dB},
\end{IEEEeqnarray*}
where the BER is the bit error rate, and $\erfc(.)$ is the complementary error function.
The Q-factor of the NNs is compared with that of DBP and linear equalization. The DBP  replaces the CD 
compensation unit at the beginning of the DSP chain and is applied with  single step per span, and 2 samples per symbol. 
This comparison is done to evaluate the effectiveness of the NN in jointly mitigating the residual CD and Kerr nonlinearity.

Fig.~\ref{fig:perf-uq}(a) shows the Q-factor gain of the unquantized \convfc\ model over the linear DSP in the TWC
fiber experiment ($K=M=40$) \cite{darweesh2022ecoc}.
The results demonstrates that the NN offers a Q-factor enhancement of $0.5$ dB at -2 dBm,
and $2.3$ dB at 2 dBm. 
The raw data before the linear DSP were not available to add the DBP curve to Fig.~\ref{fig:perf-uq}(a).
The TWC fiber link is short.
On the other hand, the nonlinearities are stronger in the fiber link in the SMF experiment than in the TWC fiber experiment, due to the longer
length. 
For the SMF  experiment, 
Fig.~\ref{fig:perf-uq}(b) shows that the \convfc\ model provides a performance similar to that of DBP with 1
sample/symbol (\sps). 
The improvement results from the mitigation of the dual-polarization nonlinearities, as well as the 
equipment's distortions. 
The \bilstm\ based receiver in the LEAF experiment (with $n_h=100$, $M=40$) also gives a comparable performance 
to the DBP as shown in Fig.~\ref{fig:perf-uq}(c).

In general, the implementation of the NN can be computationally expensive.
In order to reduce the complexity, in the next section, we quantize the NNs, 
casting the weights and activations into low precision numbers.

%%%%%%%%%%%%%%%%%%%%%%%%%%%%%%%%%%%%
%%%%% SECTION IV: Quantization of neural networks
%%%%%%%%%%%%%%%%%%%%%%%%%%%%%%%%%%%

\section{Quantization of the Neural Networks}
\label{sec:quant}

The parameters (weights and biases) of the NN, activations and input data are initially real numbers 
represented in FP 32 (FP32) or 64 bit numbers, described, \eg\ in the IEEE 754 standards. 
The implementation of the NNs in memory or computationally restricted environments  requires that these numbers 
to  represented by fewer number of bits and in different format, \eg\ in INT8. 

Define the quantization grid $\mathcal W$ as a finite set of numbers
\begin{equation*}
\mathcal W= \bigl\{\hat{w}_0, \hat{w}_1, \cdots, \hat{w}_n\bigr\},
\end{equation*}
where $\hat w_i\in\mathbb R$ are the quantization symbols. 
A continuous random variable $w\in \mathbb R$ drawn from a probability distribution $p(w)$ is quantized to $\hat w = Q(w)$, where $Q:\mathbb R\mapsto \mathcal W$ is the
quantization rule or  quantizer 
\begin{IEEEeqnarray*}{rCl}
Q(w) = \sum\limits_{i=0}^{N} \hat{w}_i \one_{I_i}(w).
\end{IEEEeqnarray*}
Here, $ I_i = [\Delta_{i},\Delta_{i+1})$, 
where $\{\Delta_{i}\}_{i=0}^{N+1}$ are the quantization thresholds, and  $\one$ is the indicator function, \ie\ 
$\one_{I_i}(w)=1$ if $w\in I_i$, and  $\one_{I_i}(w)=0$ otherwise. 
The intervals $\{I_i\}_{i=0}^N$ are the quantization cells, partitioning the real line.  
The quantization rate of $\mathcal W$ is $b= \log_2 (N+1)$ bits, assuming that $\hat{w}_i$ are equally likely. 
The hardware support is best when $b$ is a power of two, commonly $b=8$.

The quality of reproduction 
is measured by a distortion  which is often the 
mean-square error (MSE) $D(b)= \mathbb E (w-\hat w)^2$, where the expectation $\mathbb E$ is with respect to the probability distribution
of $w$ and $Q$ (if it includes random elements).
For a fixed rate $b$, the symbols $\hat w_i$ and $\Delta_i$ (or $Q(.)$) are found to minimize the distortion $D(b)$.

\def\sfactor{0.27}
\begin{figure*}[t]
\begin{center}
\begin{tabular}{c@{~~~~~~~}c@{~~~~~~}c}
    \includegraphics[width=\sfactor\textwidth]{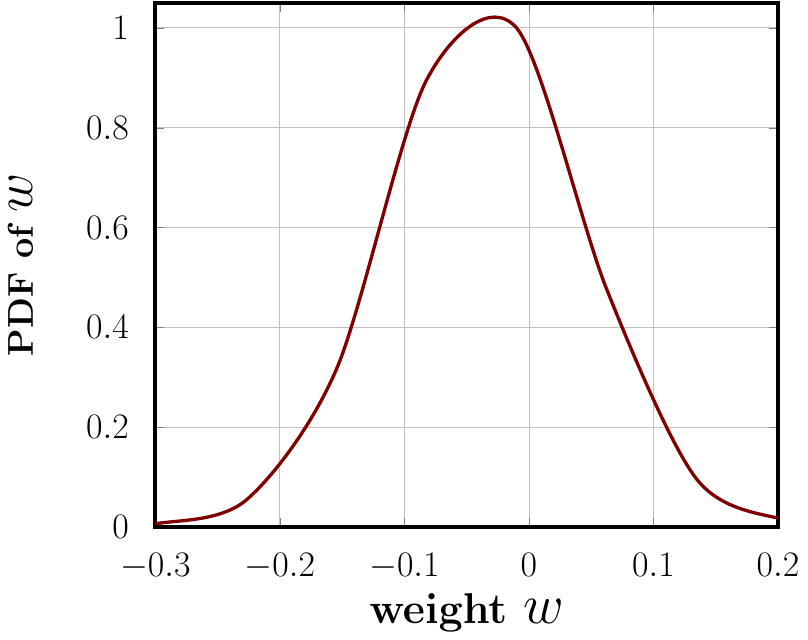}&
\includegraphics[width=\sfactor\textwidth]{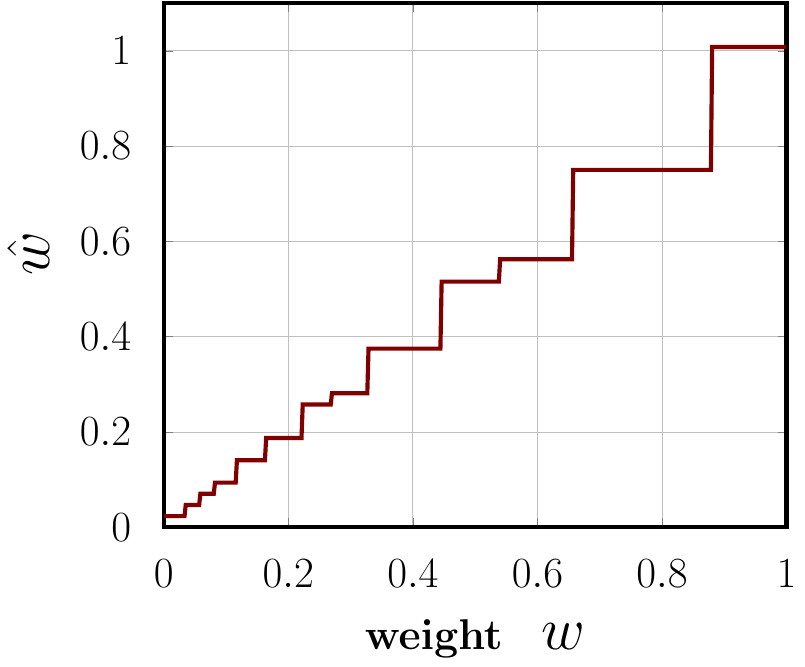}&
\includegraphics[width=\sfactor\textwidth]{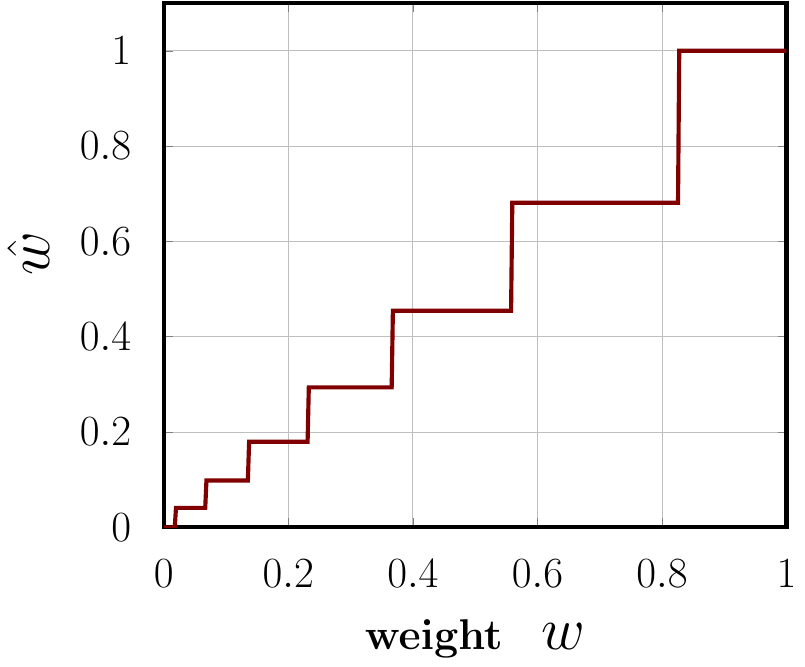}\\[2pt]
~~~~~~~{\fontsize{8}{12}\selectfont (a)} & ~~~~~~~~{\fontsize{8}{12}\selectfont  (b)} & ~~~~~~~~
{\fontsize{8}{12}\selectfont   (c)}
\end{tabular}
\end{center}
\caption{a) Probability density function (PDF) of the weights  is 
bell-shaped with non-zero mean, suggesting that uniform quantization is not optimal.
b) APoT-4, illustrating that the quantization symbols are irregularly placed; c) CP-3. }
\label{fig:w-density}
%\label{fig:qfunc}
\end{figure*}

\subsection{Quantization Schemes}

There is a significant literature on the quantization algorithms in deep learning. However, most of these algorithms 
have been developed for over-parameterized NNs with large number of parameters. These networks have many
degrees-of-freedom to compensate for the quantization error. It has been 
experimentally demonstrated that the over-parameterized NNs are rather resilient to the quantization, at least up to 8 bits.
In contrast, the NNs used for fiber equalization are small, typically with few hundred or thousands of weights, 
smaller than the models deployed even in smartphones and Internet of Things applications \cite{ignatova2019}. 
Below, we review a  number of the quantization algorithms suitable for the NN equalizers.

\subsubsection{Uniform Quantization}

In uniform quantization, the quantization symbols $\hat w_i$ are uniformly placed. 
Given a step size (or scale factor) $s$ and a zero point $z$, the uniform quantization rule is

\begin{equation*}
\hat w = s(\bar w -z),
%\Bigl\lfloor \frac{c(w,a,c)- a)}{s(a,c,N)}\Bigr\rceil s(a,c,N)+a,
\end{equation*}
where $\bar w\in \bar{\mathcal{W}}=\{0,1, \cdots, N\}$. The integer representation of $w$ is
\begin{equation*}
\bar w = \clip\left(\Bigl\lfloor \frac{w}{s}\Bigr\rceil+z; 0, N\right),
\end{equation*}
where  $\clip(w,a,b)$, $a\leq b$, is the clipping function
\begin{IEEEeqnarray*}{rCl}
\clip(w,a,b) = 
\begin{cases}
    a, & w < a,\\
    w, & a\leq w < b,\\
    b, & w\geq b,
\end{cases}
\end{IEEEeqnarray*}
in which $\lfloor x \rceil$ is the rounding function, mapping $x$ 
to an integer in $\bar{\mathcal{W}}$, \eg\ to the nearest symbol. 
The quantization grid is thus
\begin{IEEEeqnarray}{rCl}
\mathcal W_u(s,z,b) = \Bigl \{ -zs, -sz+s, \cdots, -sz+sN \Bigr\}.
\label{eq:uniform}
\end{IEEEeqnarray}

The scale factor $s$ and zero point $z$ can be determined by considering an interval $[\alpha, \beta]$ that contains 
most of the weights. Then,   $s(a,c,N){=}( \beta- \alpha)/N$ and $z=\Bigl\lfloor -\alpha/s\Bigr\rfloor$.
The interval  $[\alpha, \beta]$ is called the clipping (or clamping or dynamic) range, and is selected by a 
procedure called calibration, which may require a calibration dataset (a small set of unlabeled examples). 
The parameters of the uniform quantizer are thus $\alpha$, $\beta$, $b$ and the choice of the rounding function.

For a fixed rate $b$, the remaining parameters can be obtained by minimizing the MSE.
However, it is simpler, and sometimes about equally good (especially when $b\geq 4$), to set the clipping range to be an 
interval centered at the mean $\mu$ of $w$, with a duration proportional to the standard deviation $\sigma$ of $w$ 
\begin{IEEEeqnarray*}{rCl}
\alpha = \mu - \kappa\sigma, \quad \beta = \mu + \kappa\sigma,
\end{IEEEeqnarray*}
where, \eg\  $\kappa = 4$. Even a simpler method of calibration is setting $\alpha$ and $\beta$ to be the minimum and maximum
value of the weights $w$, respectively \cite{jacob2018quant}. The min-max choice can be sensitive to the outlier parameter values,  increasing  unnecessarily
the step size and rounding error.

In the symmetric quantization,  $z=0$. Thus, $w=0$ is mapped to $\bar w=0$ and  $\hat w=0$.
The grid of the uniform unsigned symmetric quantization is thus $\mathcal W_{\textnormal{uus}}(s) = \bigl \{ 0, s, \cdots, sN \bigr\}$. If the distribution 
of $w$ is  symmetric around the origin, symmetric signed quantization is applied, where
\begin{IEEEeqnarray}{rCl}
\mathcal W_{\textnormal{uss}}(s, b) = \Bigl \{ ks:\quad k=-(N+1)/2, \cdots,  (N-1)/2\Bigr\}.
\IEEEeqnarraynumspace
\label{eq:uniform-s}
\end{IEEEeqnarray}

The common practice is to cast the weights with the signed symmetric quantization. However, the output of the 
rectified linear unit and sigmoid activation is not symmetric.
Moreover,  the empirical distribution of the weights can sometimes be
asymmetric. For instance, Fig.~\ref{fig:w-density}
shows the weight distribution of a NN used in Section \ref{sec:results}. 
It can be seen that the distribution has a negative mean.
In these cases, asymmetric, or unsigned symmetric, quantization is used.

The quantization is said to be static if $\alpha$ and $\beta$ are known and hard-coded a priori in hardware. 
The same values are used in training and inference, and for any input. In contrast, in dynamic-range quantization, 
$\alpha$ and $ \beta$ are computed in real-time for each batch of the inputs to the NN. Since activations depend on
input, their clipping range is best determined dynamically.
This approach requires real-time computation of the statistics of the   activations, bringing about 
an overhead in computational and implementation complexity, and memory.

The computation composed of the addition and multiplication of the numbers in $\mathcal W_u$ can be performed  
with integer arithmetic, with the scale factor and zero point applied in FP32 at the end. 
In what follows, the notation UN-$b$ is used to indicate uniform quantization of the weights and activations
at  $b$ bits (with a similar notation for other quantizers).

\subsubsection{Additive Power-of-two Quantization}

In non-uniform quantization, the quantization symbols are not uniformly placed.
The hardware support for these schemes is generally limited,
due to, \eg\ the 
requirements of the iterative clustering (\eg\ via $k$-means) \cite{Shortt:06}. Thus, the majority of studies adopt uniform quantization. 
On the other hand, the empirical probability distribution of the weights is usually near bell shaped
\cite{han2015learning}; see Fig.~\ref{fig:w-density}. 
Thus, logarithmic quantization \cite{lin2016pot,li2019,elhoushi2021deepshift}
could provide lower rate for a given distortion compared 
to the uniform quantization. 

In the power-of-two (PoT) quantization, the quantization symbols are powers of two
\cite{lin2016pot}
\begin{equation*}
\mathcal W_{\textnormal{pot}} (s, r, b)=  \pm s \Bigl\{0, 2^{0}, 2^{-r}, \cdots, 2^{-r(2^{b-1}-1)} \Bigr\},
\end{equation*}
where $r\in\mathbb N$ controls the width of the distribution of symbols, and  $s\in\mathbb R$ is 
the scale factor. The scale factor is stored in FP32, but is applied after the multiply-accumulate operations, and 
can be trainable. The PoT simplifies the computation by performing the multiplications via bit shifts.
However, PoT is not flexible in the above form, and the symbols are sharply concentrated around zero.
Further, increasing the bit-width merely sub-divides the smallest quantization cell around zero, without generating 
new symbols in other cells.

The APoT introduces additional adjustable parameters, that can be used to control the
distribution of the symbols, introducing new symbols generally everywhere \cite{li2019}. 
The APoT grid  is the sum of $n$ PoT grids with a base bit-width $b_0$ and different ranges,
for a given $n\in\mathbb N$ and $b_0$. The bit-width is thus $b =nb_0$. 
Choosing $b_0$ such that $n=b/b_0$ is an integer, the quantization grid of APoT is
\begin{equation*}
\mathcal{W}_{\textnormal{apot}}(s, r, b, b_0, \gamma) {=} \pm s \sum_{i=0}^{n-1} 2^{-i}
\bigl|\mathcal{W}_{\textnormal{pot}}\bigr|(1, n,  b_0+1) +\gamma,
\end{equation*}
 where $s$ and $\gamma$ are trainable scale and shift factors in FP32, 
 the absolute value in the set $|\mathcal{W}|$ is defined per component, and $\Sigma$ is the Minkowski set sum.
It can be verified that $|\mathcal W_{\textnormal{apot}}|=2^b$. 
The shift parameter $\gamma$ allows restricting the quantized weights to unsigned numbers.

As with the PoT, the main advantage of APoT representation is that it is multiplier-free, thus 
considerably less complex than the uniform quantization. The PoT and APoT gives rise to more efficient quantizers
such as in DeepShift, where the bit-shifts or exponents are  learned directly via STE \cite{elhoushi2021deepshift}.
The use of APoT in fiber-optics equalization is discussed  in \cite{darweesh2022ecoc}.

\subsection{Companding Quantization}

In companding (CP) quantization, an appropriate nonlinear transformation is applied to the weights so that the 
distribution of the weights becomes closer to a uniform distribution, and a uniform quantizer can be applied 
afterwards \cite{yamamoto2021learnable}. 
A companding quantizer is composed of a compressor, a uniform quantizer, and an expander. The $\mu$-law is an example of a  compressor
\begin{IEEEeqnarray}{rCl}
w_c= F(w)=\sign(w)\frac{\log(1+\mu |w|)}{\log(1+\mu)},   
\label{eq:compressor}
\end{IEEEeqnarray}
where $\mu>0$ is the compression factor. Its inverse
\begin{IEEEeqnarray}{rCl}
w= \mu^{-1}\sign(w_c) \Bigl(1+\mu)^{|w_c|}-1\Bigr), 
\label{eq:expander}
\end{IEEEeqnarray}
is the expander.

Companding quantization has been widely used in data compression and digital communication. It is shown  that the 
logarithmic companding quantization can cast the weights and biases of the NN image classifiers at 2 bits \cite{peric2021robust}, and outperforms 
the uniform and APoT quantization in the same task \cite{Yamamoto_2021_CVPR}. 
However, the use of companding quantization in NN equalizers has not been investigated.

\subsection{Mixed-precision Quantization}

The majority of the quantization schemes consider fixed-precision quantization,
where a  global bit-width is predefined. 
In the mixed-precision quantization, different groups of weights or activations are 
quantized generally at different rates \cite{qu2020adaptive}.  The groups could be defined
by layers, channels, feature maps, clusters, etc.
One approach to
determine the bit-width of each group is based on the sensitivity of the model
using the Hessian matrix of the loss function \cite{NEURIPS2020_d77c7035}. 
If the Hessian matrix has a large norm on average over a particular group, 
a larger bit-width is assigned to that group.  
The output (and sometimes input) layer is often quantized at high precision, \eg\ at 16 bits, 
as it directly influences the prediction. The biases impart a small overhead and usually not quantized.
In our work, the quantization rates are 
determined from the sensitivity of the loss function.
The hardware support for mixed-precision quantization is limited compared to the fixed-precision quantization.

\subsection{PTQ and \qat}

\subsubsection{Post-training Quantization}

In PTQ, training is performed in full or half precision. The input tensor, activation outputs, 
and the weights are then quantized at fewer bits and used in inference \cite{choukroun}. In practice, the quantized values are 
stored in integer or fixed-point representations in  field-programmable gate array (FPGA) or 
application-specific integrated circuit (ASIC), and processed in arithmetic logic units with bit-wise
operations. However, the general-purpose processors include the FP processing units as well, where the numbers
are stored and processed in  FP formats. Thus, to simulate PTQ in general-purpose hardware, the quantizer $Q(.)$ 
is introduced in the computational graph of the NN after each weight, bias and activation stored in FP.

The PTQ has little overhead,
and is useful in applications where the calibration data are not available. However, quantization below 4--8 bits can cause a
significant performance degradation    \cite{hubara2020improving}. Several approaches have been proposed to
recover the accuracy in the low  bit-width regimes. 
Effort has been dedicated to  finding a smaller clipping range from the distribution of the weights, 
the layer- and channel-wise mixed precision,
and the correction of the statistical bias in the quantized parameters.
Moreover, rounding a real number to the nearest 
quantization symbol may not be optimal \cite{pmlr-v119-nagel20a}. In adaptive rounding, 
a real number is rounded to the left or right symbol based on a Bernoulli probability distribution, or deterministic optimization. 
It has been shown that PTQ-4 with adaptive rounding incurs a small loss in accuracy in some applications \cite{NEURIPS2019_c0a62e13}.

\subsubsection{Quantization-aware Training}

In \qat, quantization is co-developed with the training algorithm. This usually enhances the prediction 
accuracy of the model by accounting for the quantization error during the training.

\qat\ is  simulated by placing the quantizer function after each weight and activation in 
the computational graph of the NN.
The output of the quantizer is a piece-wise constant function of its input. This function is not 
differentiable at the points of discontinuity, and has a derivative that is zero everywhere else, \ie\
$Q'(w)= \partial \hat w/\partial w=0$.
Thus, the gradient of the loss function with respect to the weights is zero almost everywhere, 
and learning with the gradient-based methods is not directly possible.
There are a number of approaches to address the zero gradient problem, such as approximating $Q'(w)$ 
with a non-zero function, as in STE.

\qat\ usually achieves higher prediction accuracy than PTQ when quantizing at low number of bits, at the cost of the increased overhead.
On the other hand, if the approximation technique is not carefully chosen,  \qat\ may perform even worse than PTQ \cite{liu2018bi}.
Training can be performed from scratch, or from a pre-trained model, followed by  \qat\ fine-tuning the result.

\paragraph{The Straight-thorough Estimation} 
In STE, the derivative of the quantizer is approximated with the
identity function, potentially truncated on the clipping range $[\alpha, \beta]$
\begin{IEEEeqnarray}{rCl}
    Q'(w) \approx
\begin{cases}
    0, & w< \alpha,\\
    1 ,   & \alpha \leq  w < \beta,\\
    0, & w\geq \beta.
\end{cases}
\label{eq:derivative} 
\end{IEEEeqnarray}
During the NN training, in the forward pass $Q(.)$ is used. 
In the backward pass, $Q'(.)$ in 
\eqref{eq:derivative} is applied, which is then used in the chain rule to back-propagate the errors 
in training \cite{bengio2013ste,hubara2020improving}. Moreover, the weights remains in FP in the backward pass, to recover
the accuracy lost in the forward pass.
Even though (\ref{eq:derivative}) is not a good approximation to the zero, STE works surprisingly well in some models when
$b\geq 5$ 
\cite{liu2018bi}. 
The gradient is usually sensitive to quantization, even more than activations. 
It is thus either not quantized, or quantized with at least 6 bits \cite{zhou2016dorefa}.

There are non-STE approaches as well. For instance, an appropriate regularization term can be added to the loss function that penalizes 
the weights that take on values outside the quantization set. 
Another approach is the alpha-blending (AB)  quantization.

\begin{figure}[t!]
\begin{center}
\begin{tabular}{cc}
\includegraphics[width=0.24\textwidth]{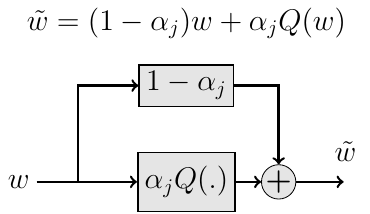} 
&
\includegraphics[width=0.22\textwidth]{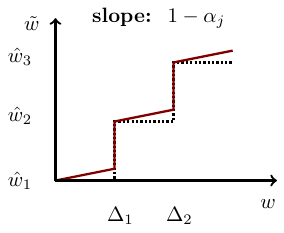}\\[2mm]
 {\fontsize{8}{12}\selectfont    (a)} & {\fontsize{8}{12}\selectfont (b)}
\end{tabular}
\end{center}
\caption{
(a) Sub-graph introduced after each weight $w$ in the computational graph of the NN in the AB quantization;
(b) the  AB quantizer, when the base quantizer is the uniform one.
}
\label{fig:AB-diagram}
\end{figure}

\paragraph{Alpha-blending Quantization}
The AB quantization addresses the problem of the quantizer's zero derivative by replacing each 
weight with a
convex combination of the full precision weight $w\in\mathbb R$ and its quantized version $\hat w = Q(w)$ \cite{liu2019learning}: 
\begin{IEEEeqnarray}{rCl}
\tilde w = (1-\alpha_j) w+\alpha_j \hat w,
\label{eq:AB-update}
\end{IEEEeqnarray}
where the coefficient $\alpha_j$ is changed from $0$ to $1$ with the epoch index $j \in \{k_{1},\cdots, k_{2}\}$ according to
\begin{IEEEeqnarray}{rCl}
\alpha_j= 
\begin{cases}
      0, & j\leq k_{1},\\
      \Bigl(\frac{k_{1}-j}{k_{2}-k_{1}}\Bigr)^{3}, & k_{1}< j\leq k_{2},\\
      1, &  j\ge k_{2},\\
\end{cases}
\label{eq:alpha}
\end{IEEEeqnarray}
for some $k_1\leq k_2$.
This approach enables a
smooth transition from the unquantized weights corresponding to $\alpha_{k_1}=0$ to
the quantized ones corresponding to $\alpha_{k_2}=1$. 
The AB quantization 
is integrated into the computational graph of the NN, by placing the sub-graph shown in Fig.~\ref{fig:AB-diagram} at the
end of  each scalar weight.

Considering $Q'(.)=0$, we have $\partial \tilde w/\partial w = 1-\alpha$, and 
$\partial L(\tilde w)/\partial w= L'\left(\tilde w\right) (1-\alpha)\neq 0$.
Thus, even though the quantizer has zero derivative, 
the derivative of the loss function with respect to $w$ is non-zero, 
and the weights are updated in the gradient-based training.
The activations can still be quantized with STE.

The AB \qat\ starts with $j=k_1$, and trains with one or more epochs. Then, $j$ 
is incremented to $k_1+1$, and the training continues, initialized with the weights obtained at $j=k_1$.
It has been shown that the AB quantization provides an improvement over 
\qat-STE in different scenarios \cite{liu2019learning}. 

Given a base quantizer $Q(.)$, the AB quantization may be viewed    as using the quantizer 
$Q_{ab}(w) =(1-\alpha_j) w+\alpha_j Q(w)$. As shown in Fig.~\ref{fig:AB-diagram}(b), when $Q(.)$ is the uniform 
quantizer, $Q_{ab}(.)$ is a piece-wise linear approximation to $Q(.)$, with slope $1-\alpha_j$. 
As $\alpha_j\rightarrow 1$, the approximation error tends to zero, and $w$ is quantized.

\subsubsection{Successive Post-training Quantization}

Successive PTQ (SPTQ) may be viewed as a combination of PTQ and \qat\ \cite{zhou2017incremental}, 
and is  particularly 
effective for quantizing small NNs such as those encountered in optical  fiber communication as discussed in \cite{darweesh2023oecc}.
The idea is to compensate for the quantization error in the training. 
The parameters 
of the NN are partitioned into several sets and sequentially quantized based on a 
PTQ scheme. 
This approach is simple and tends to perform well in practice, with a good PTQ scheme and hyper-parameter optimization.

At stage $i$,  the set of weights in the layer $\ell$ denoted by 
$\mathcal{W}^{(l)}_i$
is partitioned into two subsets $ \mathcal{W}_{i,1}^{(\ell)}$ and $ \mathcal{W}_{i,2}^{(\ell)}$ corresponding to 
the quantized and unquantized weights, respectively, \ie\ 
\begin{IEEEeqnarray}{rCl}
\mathcal{W}^{(\ell)}_i = \Bigl\{\mathcal W_{i,1}^{(\ell)}, \mathcal W_{i,2}^{(\ell)}\Bigr\},\quad 
  \mathcal{W}_{i,1}^{(\ell)}\cap \mathcal W_{i,2}^{(\ell)} = \emptyset.
\label{eq:partition}
\end{IEEEeqnarray}

The model is first trained over weights in $\mathcal{W}_{i}^{(\ell)}$ in FP32. 
Then, the resulting weights  in $\mathcal{W}_{i,1}^{(\ell)}$ are quantized under a suitable PTQ scheme. 
Next,  the weights in $ \mathcal{W}_{i,1}^{(\ell)}$ are fixed, and the model is retrained by minimizing 
the loss function with respect to the weights in $\mathcal{W}_{i,2}^{(\ell)}$, starting from the previously trained values. 
The second group is retrained in order to compensate for the quantization error arising from the first group, and make up
for the loss in the accuracy. 
In stage $i+1$, the above steps are repeated upon substitution $\mathcal W_{i+1}^{(\ell)} \stackrel{\Delta}{=} \mathcal
W_{i,2}^{(\ell)}$. The weight partitioning, group-wise quantization, and retraining is repeated until the network is fully quantized. 
The total number of partition sets is denoted by $N_p$.

In another version of this algorithm, the partitioning for all stages is set initially.
That is to say, the weights of layer $\ell$ are partitioned into $N_p$ groups $\{\mathcal{W}_{i}^{(\ell)}\}_{i=1}^{N_p}$ and successively quantized, such that at each stage the
weights of the previous groups are quantized and fixed, and those of the remaining groups are retrained.

The hyper-parameters of the SPTQ are the choice of the quantizer function in PTQ and the partitioning scheme.
There are several options for the partitioning, such as random grouping, neuron grouping and local clustering. 
It has been demonstrated that models trained with SPTQ provide classification accuracies comparable 
to their baseline counterparts trained in 32-bit, with  fewer bits \cite{zhou2017incremental}. 
Fig.~\ref{fig:SAB}(c) shows that SPTQ improves the Q-factor considerably, around 0.8 dB.

\begin{algorithm}[t]
\caption{\sab\ quantization algorithm}\label{alg:cap}
\textbf{Input: } The weights $\mathcal{W}^{(l)}$ of the layer $l$, trained in full precision;
and a quantizer $Q(.)$
\\\textbf{Output: }The low precision weights $\hat{\mathcal{W}}^{(l)}$ 
\begin{algorithmic}
    \State Initialize $\mathcal W^{(l)}_{1} = \mathcal W^{(l)}$ and  $i=1$. 

\While{$\mathcal{W}_{i}^{(l)}\neq\emptyset$}

\State Partition  $\mathcal W_{i}^{(l)}$ into $\mathcal W_{i,1}^{(l)}$ and   $\mathcal W_{i,2}^{(l)}$  

\For{$j \in \{ k_{1}, \cdots, k_{2}\}$}
\State 
For each $w\in\mathcal{W}_{i,1}^{(l)}$, calculate $\alpha_j$, and update:
\begin{IEEEeqnarray*}{rCl}
w \gets (1-\alpha_j) w +\alpha_j Q(w)
\end{IEEEeqnarray*}

\State  Fix $\mathcal{W}_{i,1}^{(l)}$, and for each $w\in\mathcal{W}_{i,2}^{(l)}$, update:
\begin{IEEEeqnarray*}{rCl}
w\gets \textnormal{~weight upon training over~}\mathcal{W}_{i,2}^{(l)}
\end{IEEEeqnarray*}

 \EndFor
\State $\mathcal W_{i+1}^{(l)} \gets \mathcal{W}_{i,2}^{(l)}$
\State $i\gets i+1$
\EndWhile
\State $\hat{\mathcal{W}}^{(l)} \gets \mathcal{W}^{(l)}_{i,1}$.
\end{algorithmic}
\end{algorithm}

\subsubsection{Successive Alpha-blending Quantization}

In this section,  we  propose \sab, a quantization algorithm suitable for the conversion of a  small full-precision
model to a low-precision one, in the low bit-width regime 1--3 bits, depending on whether or not the activations are 
quantized.

\sab\ is an iterative  algorithm with several stages, blending SPTQ and AB quantization in a particular manner 
described below.   At  stage $i$, the weights are partitioned into the set $\mathcal{W}^{(\ell)}_{i,1}$ 
and $\mathcal{W}^{(\ell)}_{i,2}$
as in \eqref{eq:partition}.
First, each weight $w\in \mathcal W_{i,1}^{(\ell)}$ is updated 
according to the AB relation  (\ref{eq:AB-update}) as 
$\tilde w = (1-\alpha_j)w+ \alpha_j \hat{w}$, where $\alpha_j$ is given by \eqref{eq:alpha} at $j=k_1$. 
Then, the weights  $\tilde w\in \mathcal W_{i,1}^{(\ell)}$ are  fixed, while those in $\mathcal W_{i,2}^{(\ell)}$ are
retrained from their previous  values. 
Next,  $\alpha_j$ is incremented  to the value in the sequence  \eqref{eq:alpha}
at $j=k_1+1$.
The process of partitioning, AB updating, and retraining is repeated until $\alpha_j=1$ is reached at $j=k_2$, 
where all weights in  $\mathcal W_{i,1}^{(\ell)}$ are
fully quantized. The algorithm then advances to the next stage $i+1$, by partitioning   $\mathcal W_{i,2}^{(\ell)}$
into two complementary sets.
The last partition is trained with the AB algorithm instead of being
fixed, to address the problem of the performance drop in the last set that was encountered in SPTQ.
The quantization process is summarized  in Algorithm \ref{alg:cap}.

Note that \sab\ is not directly a combination of SPTQ and AB: the successive retraining
strategy is distributed within  the AB algorithm with respect to $\alpha_j$. Therefore,
\sab\ quantization improves upon SPTQ and AB quantization, since each partition is not quantized in one
shot, rather is incrementally quantized by increasing $\alpha_j$. This  allows the trained set $\mathcal W_{i,2}^{(\ell)}$ to
adapt to the changes in $\mathcal W_{i,1}^{(\ell)}$.
Instead of fixing the last partition as in the SPTQ scheme, the AB algorithm is applied to train the last partition
and fix the quantization error. This modification leads to a reduction in the drop in performance occurred in the last
partition.

In uniform SAB quantization,  the grid is \eqref{eq:uniform}.
On the other hand, in the companding SAB quantization, first the compressor \eqref{eq:compressor} is applied so that the probability distribution of the weights is
approximately uniform on the clipping range. Then, all weights are quantized with the uniform \sab\ algorithm, and passed through the expander
\eqref{eq:expander}.

\subsection{Computational Complexity of the Quantized NNs}

In this Section, we present expressions for the computational complexity of the  two NN equalizers
described in Section \ref{sec:two-nns} after quantization, in order to quantify the gains of quantization 
in memory and computation.
The complexity is measured in the number of the 
elementary bit-wise operations (BO) \cite{baskin2021uniq}.
The reduction in memory is simply $1-b/32$, where $b$ is the quantization rate.

\subsubsection{FC Layers}
Consider a FC layer with $n_i$ inputs each with bit-width $b_{i}$, $n_o$
neurons at output, and per-weight bit-width of $b_w$.
There are $n_o$ inner products, each between vectors of length $n_i$. 
The main step is the BO to compute an inner product, which is bounded in Appendix~\ref{app:bo}.
From \eqref{eq:bo-inner},
\begin{IEEEeqnarray}{rCl}
    \text{BO}_{\text{FC}} \leq n_{o}\Bigl(n_{i} b_{i}b_{w}+(n_{i}-1)(b_{i}+b_{w}+\log_2(n_{i}))\Bigr).
    \IEEEeqnarraynumspace
\label{eq:bo-fc}
\end{IEEEeqnarray}

\subsubsection{Convolution Layers}

Consider a one-dimensional convolutional layer, with an input of length $n_i$ and per-element bit-width $b_{i}$, 
and a filter with length  $n_w$ and per-element bit-width $b_w$. It is assumed that the filter is padded with zeros on the boundaries
so that the number of  output features equals to the length of the input vector $n_i$ ("same padding").
This layer requires $n_i$ inner products between vectors of length $n_w$. The BO is thus   
\begin{IEEEeqnarray}{rCl}
\text{BO}_{\text{Conv}} \leq  n_{i} \Bigl(n_w b_{i}b_{w}+(n_w-1)(b_{i}+b_{w}+\log_2(n_w)\Bigr).
\IEEEeqnarraynumspace
\label{eq:bo-conv}
\end{IEEEeqnarray}

\subsubsection{LSTM Cells}

Consider the LSTM cell described in \cite[Eq.~13]{shahkarami2022complexity}, 
with an  input of length $n_i$ and hidden state of size $n_h$ at each time step. 
The cell has four augmented dense matrices with dimension $n_h\times (n_i+n_h+1)$, in the three gates 
and the cell activation state.
Suppose that the activations, and thus the hidden state, are quantized at $b_a$ bits.
The bit-width of the Cartesian product of the quantization grids is upper bounded by the sum of the individual
bit-widths.
Thus, from \eqref{eq:bo-fc}
\begin{IEEEeqnarray}{rCl}
    \text{BO}_{\text{LSTM}}& \leq & 4n_{h} \Bigl\{ (n_{h}+n_{i}+1)\times (b_{i}+b_{a})b_{w} + (n_{h}+n_{i})
        \nonumber\\
                           && \times \bigl(b_w+b_{i}+b_{a} +\log_2(n_{h}+n_{i}+1)\bigr)\Bigr\}.
\label{eq:bo-lstm}
\end{IEEEeqnarray}
Clearly, $\text{BO}_{\text{\bilstm}} = 2\text{BO}_{\text{LSTM}}$. 

Substituting  $b_1=b_2$ in \eqref{eq:bo-inner}, the storage and BO of the NN scale, respectively, linearly and quadratically 
with the bit-width.
Therefore, quantization from FP32 at 4 bits reduces the memory by 8X, and complexity by 64X.

The BO of the \convfc\ and \bilstmfc\ models are obtained by combining \eqref{eq:bo-fc}, \eqref{eq:bo-conv} and 
\eqref{eq:bo-lstm}.

\def\sfactor{0.3}
\begin{figure*}[t!]
\begin{center}
\begin{tabular}{c@{~~~~}c@{~~~~}c}
\includegraphics[width=\sfactor\textwidth]{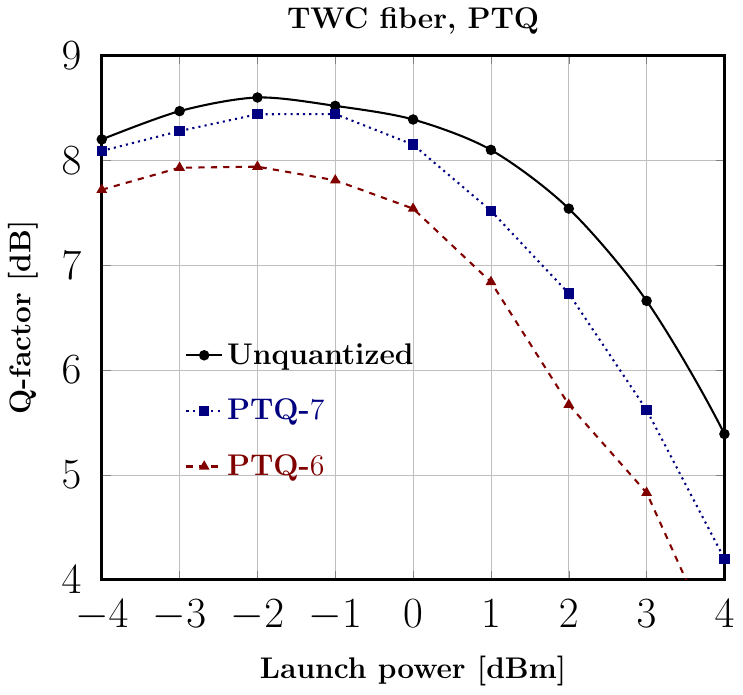} & 
\includegraphics[width=\sfactor\textwidth]{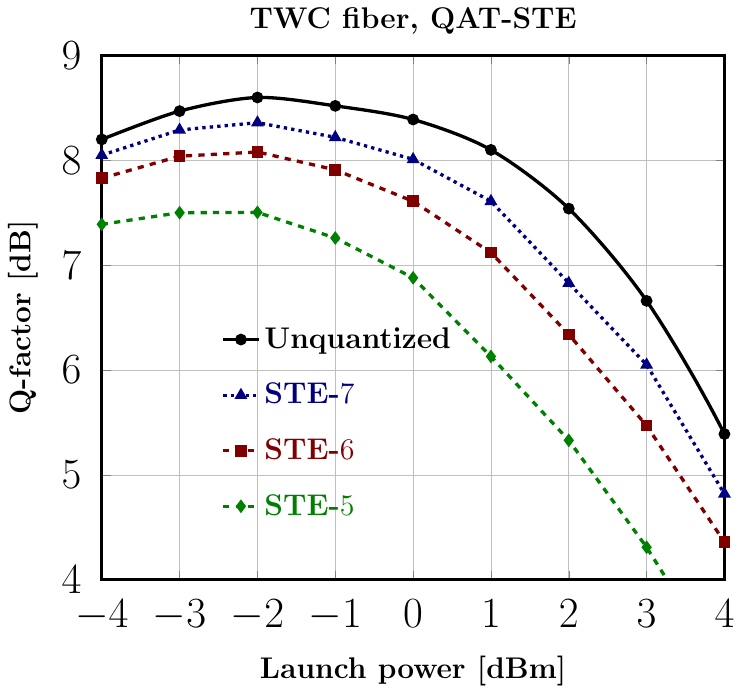} &
\includegraphics[width=\sfactor\textwidth]{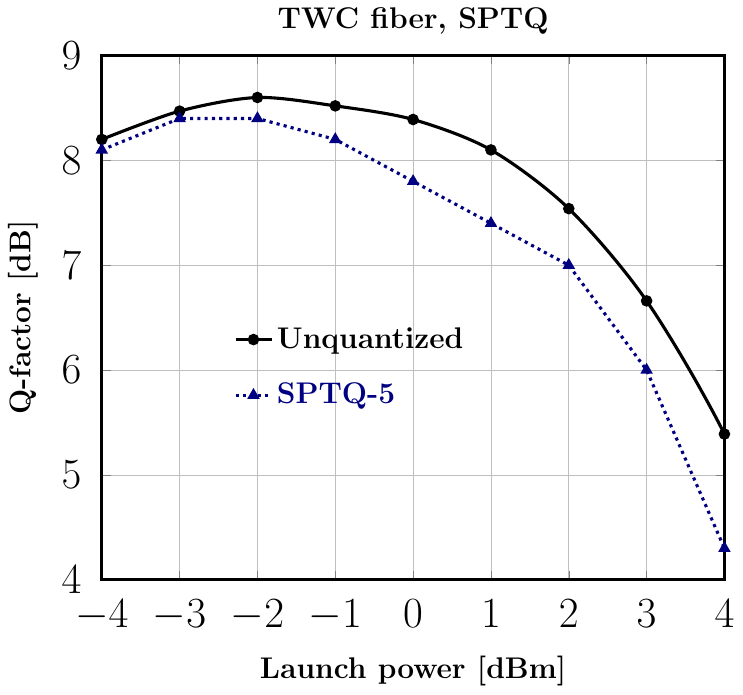} 
\\
 {\fontsize{8}{12}\selectfont (a)} & {\fontsize{8}{12}\selectfont (b)} & {\fontsize{8}{12}\selectfont (c)} 
\end{tabular}
\end{center}
\caption{Q-factor of the  NN equalizer in the TWC fiber experiment. 
a) PTQ; b)  \qat-STE; (c) SPTQ. }
\label{fig:TWC}
\end{figure*}

\subsection{Quantization of NNs in Optical Fiber Communication}
\label{sec:q-review}

The uniform  and PoT PTQ (representing fixed-point numbers) have been naturally applied when demonstrating 
the NN  equalizers in FPGA \cite{liu2021fpga,kaneda2022} or ASIC \cite{fougstedt2018asic}, usually at 8 bits.
PTQ has been applied to the NNs mitigating the nonlinear distortions in optical fiber 
\cite{he2021quant,freire2022hardware,he2021quant,fougstedt2018quant,darweesh2022ecoc,darweesh2023oecc,freire2023reducing},
and the inter-symbol interference (ISI)  in passive optical networks (PONs)   
with intensity-modulation direct-detection (IMDD) \cite{kaneda2022, huang2022low,chagnon2019quant} and in general dispersive additive
white Gaussian noise (AWGN) channels \cite{xu2019efficient}. 
In particular, the authors of \cite{kaneda2022} show that an MLP-based many-to-many equalizer outperforms the maximum likelihood 
sequence estimator in mitigating the  ISI in an IMDD 30 km PON link. 
They implement the NN in FPGA, and determine the impact of the weight resolution on the BER at 2--8 bits.
In \cite{freire2022hardware}, a multi-layer perceptron equalizing  a 1000 km SMF link is pruned and quantized with
uniform PTQ-8,  
and the reduction in BO is reported. 
The authors of \cite{fougstedt2018asic} implement the time-domain LDBP in ASIC, where the filter
coefficients, as well as the signal in each step of SSFM, are quantized.

%% PoT APoT*
The APoT is considered in  \cite{koike2021,darweesh2022ecoc, freire2023reducing}.
Fixed-point optimization-based PoT quantization is applied to an MLP equalizing an
AWGN channel in \cite{aoudia2019quant}. 
The weights are quantized  at 4  bits  and activations at 14 bits. 
The authors of \cite{koike2021} represent  the weights  using a 2-term APoT expression, for multiplier-free NN nonlinearity
mitigation in a 22x80 km SMF link. However, the quantization rate is not constrained.

The mixed-precision quantization is applied to a perturbation-based equalizer in \cite{he2021quant} (similar to the Volterra
equalizer) in a 18x100 km SMF link, in which the perturbation coefficients larger than a threshold are quantized at
large bit-width, and the rest at one bit.
Here, the quantization  also simplifies the sum expressing the equalizer, combining the 
identical or similar terms \cite{zhuge2014quant}. 

% QAT-STE
In our prior work, we compared PTQ, 
\qat-STE, APoT \cite{darweesh2022ecoc} and SPQT \cite{darweesh2023oecc} for the quantization of the NN 
 equalizers. 
However, the best rate here  is 5 bits.
The authors of \cite{freire2023reducing} study PTQ, \qat-STE and APoT, and demonstrate that the NN weights 
can be stored with a range of bit-widths and penalties, using pruning, quantization and compression.

The papers cited above mostly implement uniform, PoT, or APoT PTQ.
In our experiments, these  algorithms, and their combinations with the \qat-STE, did not achieve sufficiently small distortions in the low bit-width regime. 
The penalty due to the quantization depends on the size of the model.
The current paper addresses the quantization error, using the \sab\ algorithm that lowers the rate markedly to 1--3 bits.
Moreover, the activations are usually not quantized in the literature.
In contrast, in this paper both weights and activations are quantized.
Importantly, it will be shown in Section~\ref{sec:results} that the  quantization of activations impacts the performance
considerably.
Finally, quantization has been applied in the literature usually as an ingredient in a broader study, or combined with
pruning and compression techniques.
This paper provides a detailed analysis of the performance and complexity trade-off of different quantization
algorithms, and goes beyond the previously reported results \cite{darweesh2022ecoc,darweesh2023oecc}
in technical advances, application, and discussions.

\begin{table}[t]
\caption{\MakeUppercase{Uniform vs non-uniform quantization in TWC fiber experiment} \label{pottwc}}
\begin{center}
\begin{tabular}{ccccc}
\hline
 \multicolumn{2}{c}{Bit-width}& 
&
\multicolumn{2}{c}{Q-factor} \\  
 Convolutional& Dense &  Quantizer  & -2 dBm & 2 dBm      \\ \hline
 32&32&Unquantized&8.6&7.54\\
 6&8&Uniform&8.1&6.34\\
 6&8&ApoT&8.4&7.4\\
 \hline
\end{tabular}
 \end{center}
\end{table}

%%%%%%%%%%%%%%%%%%%%%%%%%%%%%%%%%%%%
%%%%% SECTION V: Complexity gains of quantization
%%%%%%%%%%%%%%%%%%%%%%%%%%%%%%%%%%%

\section{Demonstration of the Quantization Gains in Experiments} 
\label{sec:results}

In this Section, we determine the performance and complexity trade-off of the several quantization algorithms. 
We compute the Q-factor penalty as a function of the launch power and quantization rate,  as well as 
the reduction in the memory and computational complexity, in the three transmission experiments described 
in Section \ref{sec:dual-pol-model}.

\def\sfactor{0.3}
\begin{figure*}[t!]
\begin{center}
\begin{tabular}{c@{~~~~}c@{~~~~}c}
\includegraphics[width=\sfactor\textwidth]{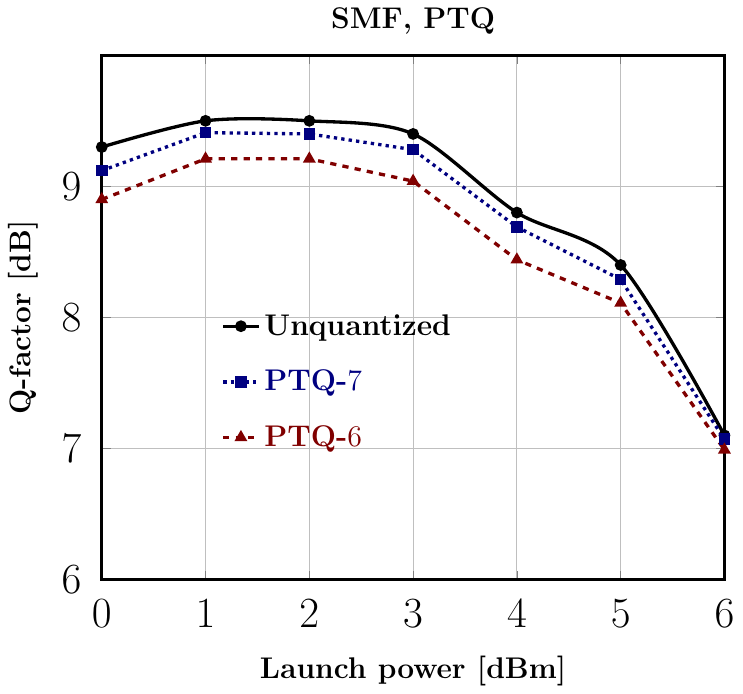} & 
\includegraphics[width=\sfactor\textwidth]{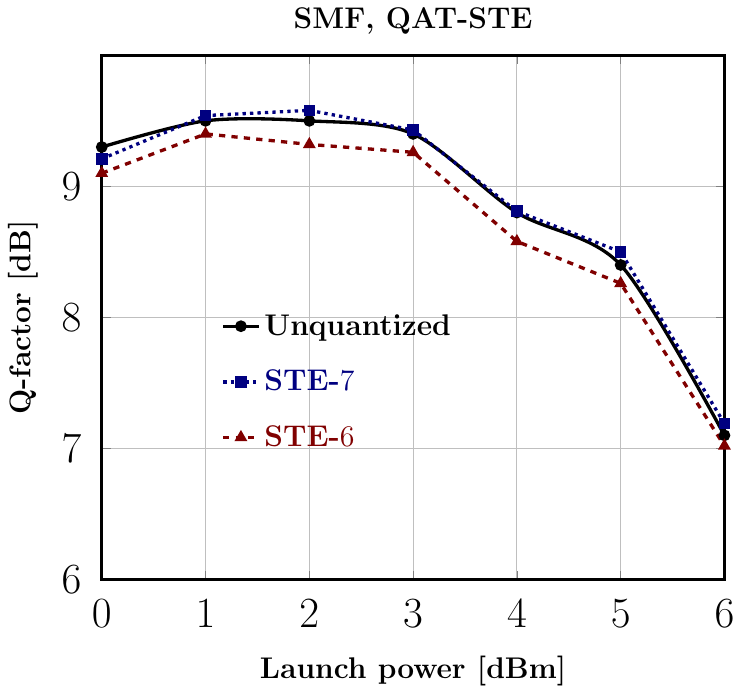} &
\includegraphics[width=\sfactor\textwidth]{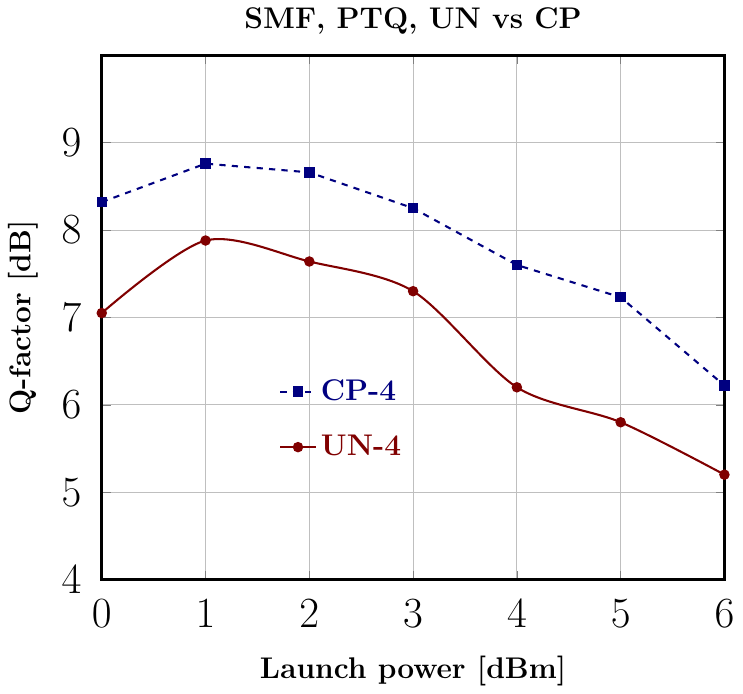} 
\\
 {\fontsize{8}{12}\selectfont (a)} &{\fontsize{8}{12}\selectfont  (b)} &{\fontsize{8}{12}\selectfont  (c)} 
\end{tabular}
\end{center}
\caption{Q-factor of the NN equalizer in the SMF experiment. 
a) PTQ; b) \qat-STE; c) uniform and companding  PTQ. }
\label{fig:SMF}
\end{figure*}

\subsection{TWC Fiber Experiment}
\label{sec:results-twc}

We consider the TWC fiber dual-polarization transmission experiment in Section~\ref{sec:TWC-exp}, with the 
\convfc\ model  in Section \ref{sec:lstm-fc}. The hyper-parameters of this model are the size of 
the convolutional filters $K$ and the number of hidden neurons $n_h$. The filters' length is set to be the residual channel 
memory, $K=M$.
This is  estimated to be $M=40$ complex symbols per polarization, through the auto-correlation function of the received 
symbols after CPE, and performance evaluation. 
The minimum number of hidden units is $n_h=100$, below which the performance rapidly drops.

The NN is trained with 600,000 symbols from a 16-QAM constellation.  A test set of 100,000 symbols is used 
to assess the performance of the NN.
Each dataset is measured at a given power, during which the BER may fluctuate in time  due to the environmental changes.
The symbols on the boundary of the data frame are eliminated to remove the
effects of the anomalies.
The NN at each power is trained and tested with  independent datasets of randomly chosen symbols at the same power.
The NN  is implemented in the Python's TensorFlow library.
The loss function is the mean-squared error, and the learning algorithm is the  Adam-Optimizer with the learning rate of 0.001.
The libraries such as TensorFlow provide functions for basic PTQ and \qat-STE, however, at 8 bits or more. 
For quantization at an arbitrary bit-width $b< 8$, the algorithms have to be directly programmed. For benchmark models in
deep learning, low bit-width implementations exist.

For quantization above 5 bits, PTQ and \qat-STE are applied, combined with APoT quantization, fixed- or  mixed precision.
In fixed-precision PTQ, the weights and activations of all layers are quantized at  6, 7 or 8 bits. 
In mixed-precision PTQ, 6  bits is assigned to the weights and activations of the 
convolutional layer, whereas the dense layer is given 8 bits due to its more significant impact on the  
performance. The Q-factor is nearly not impacted at 8 bits. 
Fig.~\ref{fig:TWC}(a) demonstrates that fixed-precision PTQ-6 incurs a penalty of 0.7 dB at -2 dBm compared to the
unquantized NN, and 1.9 dB at 2 dBm. This comes with a gain of  $81\%$ reduction in the 
memory usage and a $95\%$ reduction in the computational complexity.

The Q-factor improves using the \qat-STE, as depicted in Fig.~\ref{fig:TWC}(b). Here, the weights  are initialized with 
random values, then trained and quantized at 5, 6, and 7 bits, and the activations at 6 bits.
In this case, the drop is reduced to 0.5 dB at -2 dBm, and 1.2 dB at 2 dBm. 
As the transmission power is increased, the penalty due to the quantization  increases. 

The distribution of the weights of the dense layer is bell-shaped, as shown in  Fig.~\ref{fig:w-density}.
In consequence, assigning more quantization
symbols around the mean is a reasonable strategy.  
The APoT quantization delivers a good performance, with a Q-factor penalty of less than
$0.2$ dB at $-2$ and $2$ dBm, as seen in Table~\ref{pottwc}.

%%%%%%%%%%%%%%%%%%%%%%%%%%%%%%%%%%%%%%% SPTQ %%%%%%%%%%%%%%%%%%%%%%%%%%%%%%%%%%%%%%%% 

The uniform SPTQ is applied, by assigning 5 bits to the weights and activations of the dense layer.
The convolutional layer is given 8 bits, but this layer has few weights, and little impact on the complexity.
Fig.~\ref{fig:TWC}(c) shows that SPTQ at 5 bits leads to 0.2 dB Q-factor drop at -2 dBm, 
and 0.5 dB at 2 dBm.
It can be seen that SPTQ  outperforms the more complex \qat-STE by 2 bits at the same power \cite{darweesh2023oecc}.
Fig.~\ref{fig:SAB}(c) shows that increasing the partition size can notably enhance the Q-factor.
Similar conclusions are drawn for SPTQ-4, as seen in Table~\ref{tab:TWC-SPTQ}.

%%%%%%%%%%%%%%%%%%%%%%%%%%%%%%%%%%%%%% SAB %%%%%%%%%%%%%%%%%%%%%%%%%%%%%%%%%%%%%%%%%%%%%

For quantization below 5 bits, we apply \sab. 
In a first study, we consider fixed-precision quantization, where the weights and activations are quantized at 4 bits
successively over 4 partitions. The results in Table \ref{tab:sab-4} indicate that \sab\ outperforms SPTQ and AB,  
with a performance drop of  0.5~dB near optimal power. 
In contrast, SPTQ and AB quantization resulted in a 1.2 dB drop in performance. 
In a second study, we apply mixed-precision \sab, giving more bits to the last partition. We consider a partition of size 4 with 
the weights and activations in the first three partition sets quantized at 4 bits, and in the last  set at 6 bits, 
averaging to 4.5 bits. The results are shown 
in  Fig.~\ref{fig:SAB}(a), indicating the Q-factor drop of 0.17 dB at -2 dBm and 0.24 dB at 2 dBm.
This comes with  $86\%$ reduction in memory usage, and  $94\%$ in computational complexity.

%%%%%%%%%%%%%%%%%%%%%%%%%%%%%%%%%%%%%%%%%%%%%%%%%%%%%%%%%%%%%%%

\def\sfactor{0.3}
\begin{figure*}[t!]
\begin{center}
\begin{tabular}{c@{~~~~}c@{~~~~}c}
\includegraphics[width=\sfactor\textwidth]{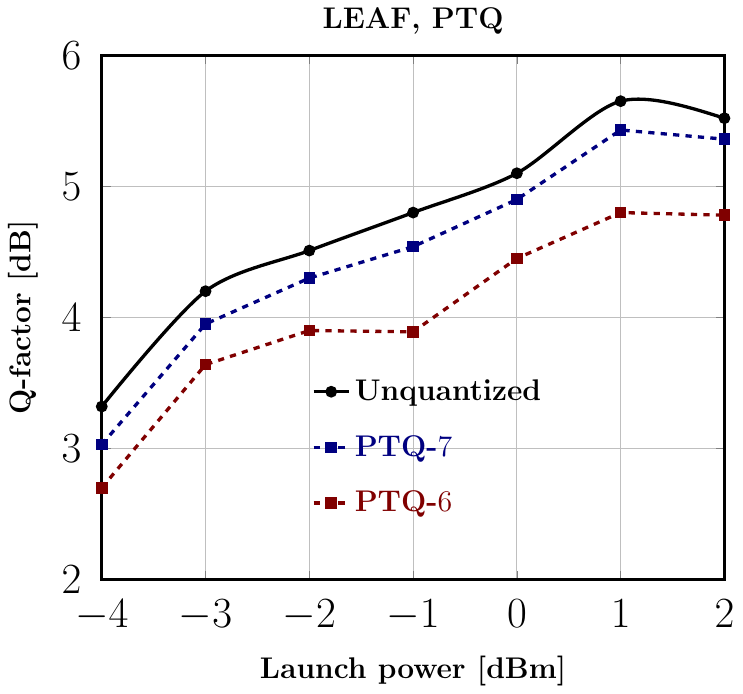} & 
\includegraphics[width=\sfactor\textwidth]{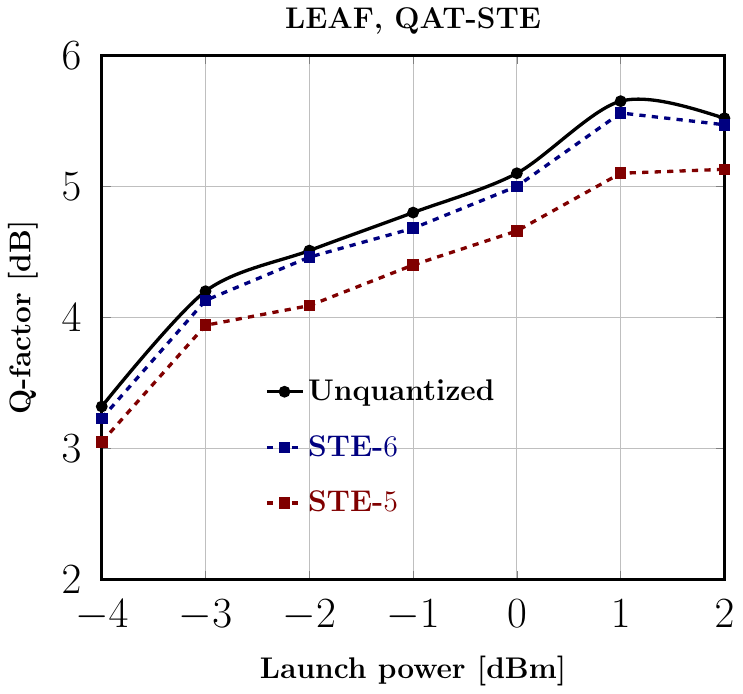} & 
\includegraphics[width=\sfactor\textwidth]{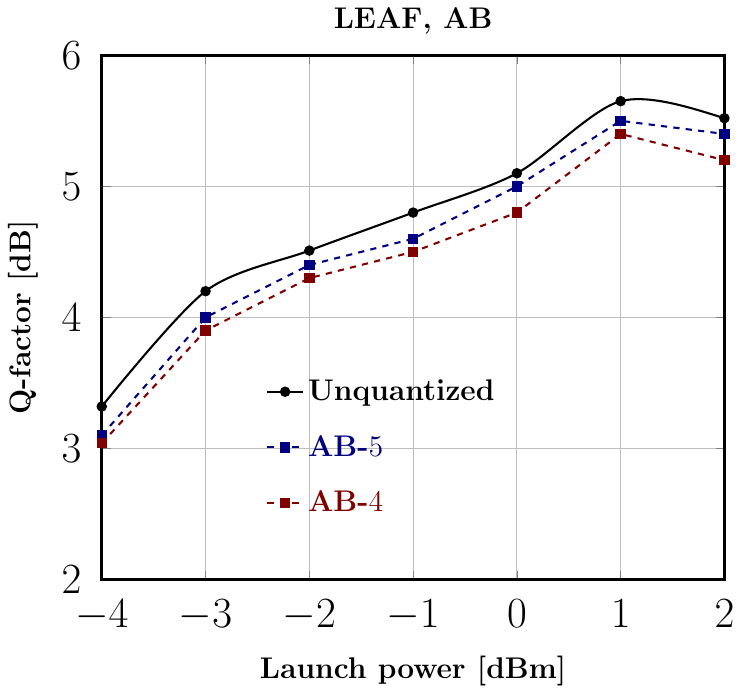}
\\
 {\fontsize{8}{12}\selectfont  (a)} &{\fontsize{8}{12}\selectfont  (b)} &{\fontsize{8}{12}\selectfont  (c)} 
\end{tabular}
\end{center}
\caption{Q-factor of the NN equalizer in the LEAF experiment. 
a PTQ; b)  \qat-STE; (c) AB quantization. }
\label{fig:LEAF}
\end{figure*}

\subsection{SMF  Experiment}

\begin{table}[t]
\caption{ \MakeUppercase{Q-factor of SPTQ-4, TWC Fiber Experiment}\label{tab:TWC-SPTQ}}
\centering
\begin{tabular}{c|cccccccc} 
\hline
 $N_p$&\multicolumn{8}{c}{Q-factor}\\
 &$\mathcal W_{1}$&$\mathcal W_{2}$&$\mathcal W_{3}$&$\mathcal W_{4}$&$\mathcal W_{5}$
 &$\mathcal W_{6}$&$\mathcal W_{7}$&$\mathcal W_{8}$\\
 \hline
 2&7.13&\textbf{5.6}  & & & & & &  \\
 4&7.5&7.33&7.33&\textbf{6.3}& & & &  \\
 8&7.56&7.5&7.4&7.33&7.33&7.33&7.33&\textbf{6.6} \\
\end{tabular}
\end{table}

We consider the SMF experiment described in Section~\ref{sec:SMF-exp}, with the \convfc\ model.
The NN parameters and the quantization algorithms are similar to those in the TWC fiber experiment.

For quantization above 5 bits, PTQ-6 led to a Q-factor drop of 0.3 dB at 1 dBm, and 0.4 dB at 4 dBm,  as shown in Fig.~\ref{fig:SMF} (a). 
For \qat-STE-6, as shown in Fig.~\ref{fig:SMF}(b), the drop is 0.1 dB at 1 dBm, and 0.2 dB at 4 dBm.

For quantization below 5 bits, first the companding PTQ is applied. 
Fig.~\ref{fig:SMF}(c) shows that this quantizer outperforms the uniform quantization at 4 bits by about a dB, due to the non-uniform 
distribution of the weights of the dense layer.  
It is found that, while the APoT works well in the large bit-width regime $b\geq 6$ (as in the TWC fiber experiment), 
it is uncompetitive at low bit-widths.

%%%%%%%%%%%%%%%%%%%%%%%%%%%%%%%%%%%%%%%%%%%%%%%%%%% SAB %%%%%%%%%%%%%%%%%%%%%%

Next, we apply \sab\ quantization, in a partition of size 4, where the weights  in the first 3 sets are quantized 
at 3 bits, and in the last set at 6 bits, with the average rate of 3.75 bits. 
The activations for all partition sets are quantized at 3 bits. The uniform and companding versions are both studied.
Fig.~\ref{fig:SAB}(b) shows the results.    
Uniform \sab\ quantization results in a Q-factor drop of 0.3 dB at 1 dBm, and 0.6 dB at 4 dBm. 
This quantizer offers a reduction in memory usage and computational complexity, by $88\%$ and $94\%$, respectively.
Applying the companding \sab\ quantization, the Q-factor drop is reduced to 0.2 dB at 1 dBm.

\subsection{LEAF Experiment}

The NN in this experiment is the \bilstmfc\ equalizer, described in 
Section~\ref{sec:lstm-fc}. There are $n_h=100$ hidden neurons, and the input size is $\bar n_i=4(M+1)$, $M=40$.
This model is found to be prone to the quantization error, because small errors can be amplified 
by the internal activations, and accumulate over long input temporal sequences.
Thus, we quantize the weights and biases of the forget, input and output gates, 
as well as the activations at the output of the cell.
However, the internal activations remain in  full precision.

Fig.~\ref{fig:LEAF} (a) shows that PTQ-6 incurs a Q-factor penalty of $0.9$ dB at 1 dBm,  
and $1.2$ dB at $-1$ dBm, respectively, while lowering the computational 
complexity by  $79\%$ and the memory usage  by $81\%$.
\qat-STE significantly improves the Q-factor, as shown in Fig.~\ref{fig:LEAF} (b).
At 6 bits, the drop is  $0.1$ dB at 1 dBm, and $0.4$ dB at $-1$ dBm.
At 5 bits, the penalty is $0.3$ dB at both $1$ dBm and $-1$ dBm, with $82\%$ reduction in computational complexity 
and $84\%$ in memory usage.

Fig.~\ref{fig:LEAF}(c) shows that the AB quantizer at 4 and 5 bits outperforms PTQ and \qat\
Specifically, the Q-factor drop is only $0.2$ dB at -1 dBm, and $0.15$ dB at 1dBm.

\begin{table}[t]
 \caption{\MakeUppercase{Fixed-precision quantization, TWC fiber experiment\label{tab:sab-4}}}
\begin{center}
\begin{tabular}{c|cc}
\hline
 Quantization scheme&bit-width&Q-factor\\
 \hline
 Unquantized&32&7.5\\
SPTQ&4&6.3\\
   AB&4&6.3\\
    \sab &4&7.0\\
\end{tabular}
 \end{center}
\end{table}

\subsection{Quantization of the Weights, but not Activations}
\label{sec:binary}

In the previous sections, the weights  and activations were both quantized. It can be seen that there is a cut-off bit-width around 5--6 bits, 
below which the performance of the \qat-STE rapidly drops. 
Upon investigation, we noticed that the quantization of the activations substantially impacts the Q-factor.
The activation functions are nonlinear, and could amplify the quantization error. 
In this section, we consider 
quantizing the weights of the NN but not activations. The bit-width of the activations can still be reduced from 32 to 
8 with negligible performance drop. Therefore, the activations are  quantized, at 8 bits.

In a first study, we quantize the weights of the \convfc\ model in the SMF experiment, using the fixed-precision 
\sab\ algorithm with a partition of size 4.
The results are included in Table~\ref{tabb}, showing that the Q-factor drop at the optimal power is
minimal, when the dense layer is  quantized at as low as 3 bits. 
In a second study, we apply the mixed-precision \sab\ quantization with the same parameters.
The first three partitions are quantized at 1 bit, and the last one at 4 bits.
We obtain a quantization rate of 1.75 bits/weight, with 0.6 dB degradation in Q-factor, outperforming 
the state-of-the-art using the \qat-STE w/wo APoT by 2 dB.
This important result demonstrates that low-complexity nearly-binary NNs 
can mitigate nonlinearities  in optical fiber communication.

\def\sfactor{0.3}
\begin{figure*}[t!]
\begin{center}
\begin{tabular}{c@{~~~~}c@{~~~~}c}
\includegraphics[width=\sfactor\textwidth]{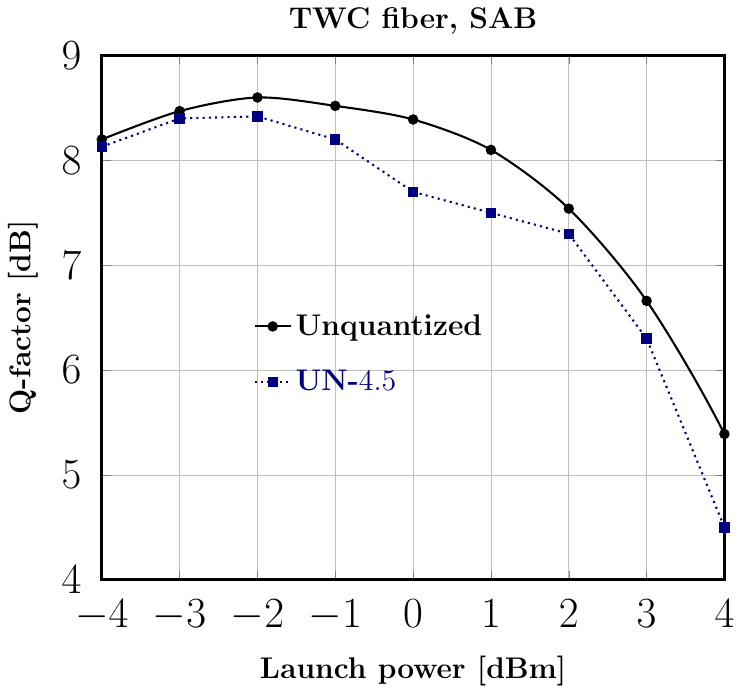}& 
\includegraphics[width=\sfactor\textwidth]{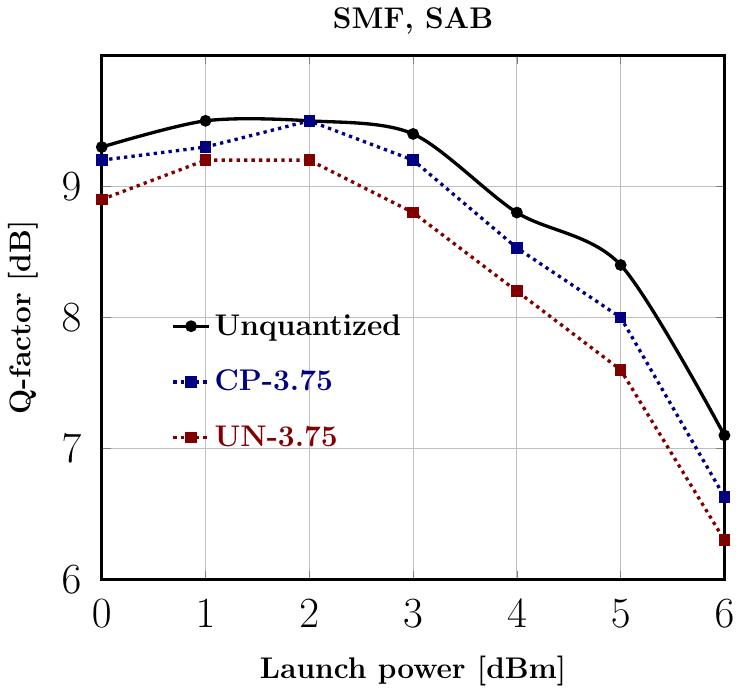} &
\includegraphics[width=\sfactor\textwidth]{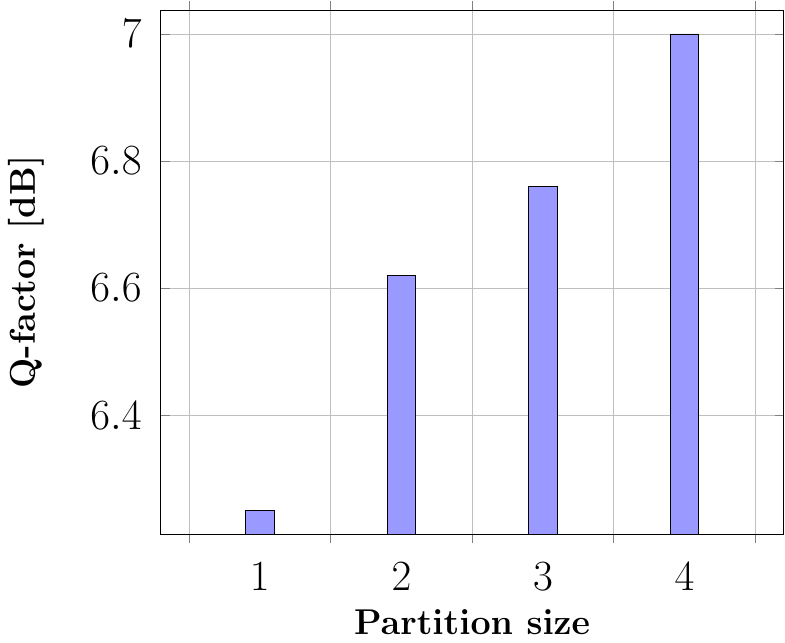}  
\\
~~~~{\fontsize{8}{12}\selectfont (a)} & ~~~~~{\fontsize{8}{12}\selectfont (b)}  & ~~~~~~~~{\fontsize{8}{12}\selectfont
    (c)} 
\end{tabular}
\end{center}
\caption{Q-factor of the NN equalizer with \sab\ quantization in 
a) TWC fiber and b) SMF experiment. c) Impact of the partition size in SPTQ-5 in TWC fiber experiment. 
}
\label{fig:SAB}
\end{figure*}

In the so called ``extreme quantization,'' the NNs are quantized at 1 or 2 bits 
\cite{courbariaux2015bc,courbariaux2016binarized, darabi2020regularized, liu2022ternary,zhou2018dorefanet}.
Many approaches to the binary and ternary NNs have been proposed, \eg\ based on better approximations 
to the derivative of the quantizer than in the STE.
However,  we tested some of these approaches in our experiments, 
and did not observe notable gains over the linear equalization. 
Consequently, while extreme quantization has shown success in large models in computer vision, further work is
needed to determine if it can be adapted and successfully applied to the small NN equalizers in optical fiber
communication.

%%%%%%%%%%%%%%%%%%%%%%%%%%%%%%%%%%%%
%%%%% SECTION VI: Conclusions
%%%%%%%%%%%%%%%%%%%%%%%%%%%%%%%%%%%%

\section{Conclusions}
\label{sec:conc}

The  paper shows that low-complexity quantized NNs can mitigate nonlinearities in optical fiber transmission.
The QAT-STE partially mitigates the quantization  error during the training, and 
is effective in the large bit-width regime with $b>5$ bits. 
The companding quantization improves the Q-factor of the baseline schemes considerably, especially at low bit-widths.
There is a cut-off bit-width of around 5 bits below which the penalty of the \qat-STE rapidly increases. 
In the low bit-width regime with $b\leq 5$ bits,  companding \sab\ quantization is the method of choice.
There is a considerable performance penalty due to the quantization of activations. 
The weights of the NN can be quantized at 1.75 bits/parameter with $\leq 0.5$ dB
penalty, if the activations are quantized at $b\geq 8$ bits.
The weights and  activations can be quantized at 3.75 bits/parameter, with minimal penalty.   
The LSTM-based receivers can be prone to the quantization error, due to the error amplification and propagation.
Fully binary NN equalizers remain to be studied.

%%%%%%%%%%%%%%%%%%%%%%%%%%%%%%%%%%%%
%%%%% Appendices 
%%%%%%%%%%%%%%%%%%%%%%%%%%%%%%%%%%%%

\appendices

\section{Bit-wise Operations for an Inner Product}
\label{app:bo}

The cost of computation is measured here by the required  bit-wise operations AND $\myand$, OR $\myor$, XOR $\myxor$, NOT and SHIFT \cite{baskin2021uniq}.

\subsection{Addition and Multiplication of Integers}
\label{app:bo-sum}

The sum $z = x+y$ of the integers $x$ and $y$ each with bit-width $b$ is an integer 
with bit-width $b+1$, with carry-over.
Below, we show that $z$ can be computed in $\zeta b$ BO, where $\zeta$ 
depends on the computing algorithm. 

Denote the binary representation of $x$, $y$ and $ z $ with  $x_1x_2\cdots x_b$,  $y_1y_2\cdots y_b$,
and $z_1z_2\cdots z_{b+1}$, respectively.
Let $c_1c_2\cdots c_{b+1}$ be the carry-over  binary sequence, initialized with $c_1=0$. 
Then, for $i\in \bigl\{1,2, \cdots, b+1 \bigr\}$ 
\begin{IEEEeqnarray}{rCl}
z_i= t\myxor c_i, \quad c_{i+1}= (x_i \myand y_i) \myor (t \myand c_i),
\label{eq:sum-ints}
\end{IEEEeqnarray}
where $t = x_i\myxor y_i$. 
Thus, computing $z$ using \eqref{eq:sum-ints} takes $5b$ BO, \ie\  $\zeta=5$. 
This approach requires one bit storage for $t$, and $2b$ bits transmission for memory access.

Consider the multiplication of the integers $\bar z= x y$, where $x$ has  bit-width $b_1$ and $y$ has $b_2$ bits.
Clearly, the bit-width of $\bar z$ is $b_1+b_2$.
The multiplication $2^iy$, $i\in\mathbb N$, can be performed with one BO, by shifting the $y$ in the binary form $i$ positions
to the left, and zero padding from right.
The result is a binary sequence of the maximum length $b_1+b_2$, and maximum $b_2$ non-zero bits.
Expanding $x$ as a sum of $b_1$ PoT numbers,  $\bar z$ is expressed as the sum of $b_1$ binary sequences, 
each with up to $b_2$ non-zero elements. Thus, $\textnormal{BO}=\zeta b_1 b_2$.

The value of $\zeta$ can change with the algorithm, and is immaterial. 
In this paper, we assume $\zeta=1$. 
The computation of  $z$ and $\bar z$ above may not be optimal; hence the BOs 
are upper bounds.

\subsection{The Inner Product}

The sum of $n$ numbers of bit-width $b$ can be performed in $\log_2 n$ steps 
by pairwise addition (assuming for simplicity that $n$ is a PoT number).
The sum has bit-width $b+\log_2(n)-1$ bits. The  BO can be bounded as below, or obtained
from \cite{freire2022complexity}.
\begin{IEEEeqnarray}{rCl}
    \textnormal{BO}_{\textnormal{sum}} &\leq& b \times \frac{n}{2}+  (b+1) \times \frac{n}{4}+\cdots +      (b+ \log_2(n)-1)\times 1
\nonumber\\
&=& \frac{n}{2}\Bigl[ b\sum\limits_{k=0}^{\log_2 (n)-1} 2^{-k} + \sum\limits_{k=1}^{\log_2 (n)-1} k2^{-k}\Bigr]
\nonumber\\ &\leq& \frac{n}{2}\Bigl[ (b +\log_2 n - 1)\sum\limits_{k=0}^{\log_2(n)-1} 2^{-k} \Bigr]
\nonumber 
%\\&= & \zeta n(b+\log n-1)  ( 1-\frac{1}{n})
\nonumber \\&=& 
 (b+\log_2 n) (n-1).
\label{eq:bo-sum}
\end{IEEEeqnarray}

Consider the inner product $ y= \vect w^T\vect x$, where $\vect w= (w_1, w_2, \cdots, w_n)$, $\vect x= (x_1, x_2, \cdots, x_n)$, 
and where $w_i$ and $x_i$ have, respectively, bit-width $b_1$ and $b_2$, $\forall i$.
Then, $y$ has bit-width $b_1+b_2+\log_2(n)-1$ bits.
The products $\{w_ix_i\}_{i=1}^n$ are calculated in $nb_1b_2$ BO.
Their sum is computed in BO given in \eqref{eq:bo-sum} with $b=b_1+b_2$.
Thus
\begin{IEEEeqnarray}{rCl}
\textnormal{BO}_{\textnormal{inner}} \leq nb_1b_2+ (n-1)(b_1+b_2+\log_2 n).
\label{eq:bo-inner}
\end{IEEEeqnarray}

\begin{table}[t]
\caption{\MakeUppercase{Fixed-precision \sab\ quantization, SMF experiment}\label{tabb}}
\begin{center}
\begin{tabular}{ccccc|c}
\hline
 \multicolumn{5}{c|}{Bit-width}&Q-factor\\
 $\mathcal W_{1}$&$\mathcal W_{2}$&$\mathcal W_{3}$&$\mathcal W_{4}$&Activation&\\
 \hline
 32&32&32&32&32&9.5\\
 3&3&3&3&8&9.2\\
 2&2&2&2&8&8.0\\
 1&1&1&4&8&8.9\\
\end{tabular}
 \end{center}
\end{table}

\bibliographystyle{IEEEtran}

%\bibliography{ref2}

\begin{thebibliography}{10}
\providecommand{\url}[1]{#1}
\csname url@samestyle\endcsname
\providecommand{\newblock}{\relax}
\providecommand{\bibinfo}[2]{#2}
\providecommand{\BIBentrySTDinterwordspacing}{\spaceskip=0pt\relax}
\providecommand{\BIBentryALTinterwordstretchfactor}{4}
\providecommand{\BIBentryALTinterwordspacing}{\spaceskip=\fontdimen2\font plus
\BIBentryALTinterwordstretchfactor\fontdimen3\font minus
  \fontdimen4\font\relax}
\providecommand{\BIBforeignlanguage}[2]{{%
\expandafter\ifx\csname l@#1\endcsname\relax
\typeout{** WARNING: IEEEtran.bst: No hyphenation pattern has been}%
\typeout{** loaded for the language `#1'. Using the pattern for}%
\typeout{** the default language instead.}%
\else
\language=\csname l@#1\endcsname
\fi
#2}}
\providecommand{\BIBdecl}{\relax}
\BIBdecl

\bibitem{savory2010digital}
S.~J. Savory, ``Digital coherent optical receivers: Algorithms and
  subsystems,'' \emph{{IEEE} J.\ Sel.\ Top. Quantum Electron.}, vol.~16, no.~5,
  pp. 1164--1179, {Sept./Oct.} 2010.

\bibitem{agrell2016roadmap}
E.~Agrell, M.~Karlsson, A.~Chraplyvy, D.~J. Richardson, P.~M. Krummrich,
  P.~Winzer, K.~Roberts, J.~K. Fischer, S.~J. Savory, B.~J. Eggleton
  \emph{et~al.}, ``Roadmap of optical communications,'' \emph{J.\ Opt.},
  vol.~18, no.~6, p. 063002, May 2016.

\bibitem{dar2017}
R.~Dar and P.~J. Winzer, ``Nonlinear interference mitigation: Methods and
  potential gain,'' \emph{{IEEE} J. Lightw. Technol.}, vol.~35, no.~4, pp.
  903--930, Feb. 2017.

\bibitem{gibson1989}
G.~Gibson, S.~Siu, and C.~Cowan, ``Application of multilayer perceptrons as
  adaptive channel equalisers,'' \emph{IFAC Proceedings Volumes}, vol.~23,
  no.~1, pp. 573--578, Apr. 1989.

\bibitem{ibnkahla2000}
M.~Ibnkahla, ``Applications of neural networks to digital communications -- a
  survey,'' \emph{Signal process.}, vol.~80, no.~7, pp. 1185--1215, Jul. 2000.

\bibitem{jarajreh2015}
M.~A. Jarajreh, E.~Giacoumidis, I.~Aldaya, S.~T. Le, A.~Tsokanos,
  Z.~Ghassemlooy, and N.~J. Doran, ``Artificial neural network nonlinear
  equalizer for coherent optical {OFDM},'' \emph{{IEEE} Photon. Technol.
  Lett.}, vol.~27, no.~4, pp. 387--390, Feb. 2015.

\bibitem{zhang2018}
S.~Zhang, F.~Yaman, E.~Mateo, and Y.~Inada, ``Neuron-network-based nonlinearity
  compensation algorithm,'' in \emph{Eur.\ Conf.\ Opt.\ Commun.\ Conf.}, Sep.
  2018, pp. 1--3.

\bibitem{butler2021}
R.~M. Butler, C.~Hager, H.~D. Pfister, G.~Liga, and A.~Alvarado, ``Model-based
  machine learning for joint digital backpropagation and {PMD} compensation,''
  \emph{{IEEE} J. Lightw. Technol.}, vol.~39, no.~4, pp. 949--959, Feb. 2021.

\bibitem{vanhoucke2011}
V.~Vanhoucke, A.~Senior, and M.~Z. Mao, ``Improving the speed of neural
  networks on {CPU}s,'' in \emph{Adv.\ Neural Info.\ Process.\ Syst.}, Dec.
  2011, {Deep Learning and Unsupervised Feature Learning Workshop}.

\bibitem{han2016compress}
S.~Han, H.~Mao, and W.~J. Dally, ``Deep compression: {C}ompressing deep neural
  networks with pruning, trained quantization and {H}uffman coding,'' \emph{The
  Int.\ Conf.\ Learn. Rep.}, May 2016.

\bibitem{hubara2017quant}
I.~Hubara, M.~Courbariaux, D.~Soudry, R.~El-Yaniv, and Y.~Bengio, ``Quantized
  neural networks: {T}raining neural networks with low precision weights and
  activations,'' \emph{J.\ Mach.\ Learn.\ Res.}, vol.~18, no.~1, pp.
  6869--6898, Jan. 2017.

\bibitem{jacob2018quant}
B.~Jacob, S.~Kligys, B.~Chen, M.~Zhu, M.~Tang, A.~Howard, H.~Adam, and
  D.~Kalenichenko, ``Quantization and training of neural networks for efficient
  integer-arithmetic-only inference,'' in \emph{IEEE/CVF Conf.\ Comput.\ Vision
  Pattern Recognit.}, Jun. 2018, pp. 2704--2713.

\bibitem{banner2019post}
R.~Banner, Y.~Nahshan, and D.~Soudry, ``Post training 4-bit quantization of
  convolutional networks for rapid-deployment,'' \emph{Adv.\ Neural Info.\
  Process.\ Syst.}, vol.~32, Dec. 2019.

\bibitem{courbariaux2015bc}
M.~Courbariaux, Y.~Bengio, and J.-P. David, ``Binaryconnect: {T}raining deep
  neural networks with binary weights during propagations,'' \emph{Adv.\ Neural
  Info.\ Process.\ Syst.}, vol.~28, Dec. 2015.

\bibitem{courbariaux2016binarized}
M.~Courbariaux, I.~Hubara, D.~Soudry, R.~El-Yaniv, and Y.~Bengio, ``Binarized
  neural networks: {T}raining deep neural networks with weights and activations
  constrained to +1 or -1,'' in \emph{Adv.\ Neural Info.\ Process.\ Syst.},
  vol.~29, Dec. 2016.

\bibitem{krishnamoorthi2018}
R.~Krishnamoorthi, ``Quantizing deep convolutional networks for efficient
  inference: {A} whitepaper,'' \emph{arXiv:1806.08342}, 2018.

\bibitem{hinton2012course}
G.~Hinton, ``Neural networks for machine learning,'' Coursera course, video
  lectures, 2012.

\bibitem{bengio2013ste}
Y.~Bengio, N.~L{\'e}onard, and A.~Courville, ``Estimating or propagating
  gradients through stochastic neurons for conditional computation,''
  \emph{arXiv:1308.3432}, Aug. 2013.

\bibitem{agrawal2021}
G.~P. Agrawal, \emph{Fiber-optic communication systems}, 5th~ed.\hskip 1em plus
  0.5em minus 0.4em\relax John Wiley \& Sons, 2021.

\bibitem{fatadin2009}
I.~Fatadin, D.~Ives, and S.~J. Savory, ``Blind equalization and carrier phase
  recovery in a {16-QAM} optical coherent system,'' \emph{{IEEE} J. Lightw.
  Technol.}, vol.~27, no.~15, pp. 3042--3049, May 2009.

\bibitem{pfau2009PN}
T.~Pfau and R.~No{\'e}, ``Phase-noise-tolerant two-stage carrier recovery
  concept for higher order {QAM} formats,'' \emph{{IEEE} J.\ Sel.\ Top. Quantum
  Electron.}, vol.~16, no.~5, pp. 1210--1216, { Sept.-Oct.} 2009.

\bibitem{sidelnikov2018equalization}
O.~Sidelnikov, A.~Redyuk, and S.~Sygletos, ``Equalization performance and
  complexity analysis of dynamic deep neural networks in long haul transmission
  systems,'' \emph{Opt. Exp.}, vol.~26, no.~25, pp. 32\,765--32\,776, Dec.
  2018.

\bibitem{Karanov:18}
B.~Karanov, M.~Chagnon, F.~Thouin, T.~A. Eriksson, H.~B\"{u}low, D.~Lavery,
  P.~Bayvel, and L.~Schmalen, ``End-to-end deep learning of optical fiber
  communications,'' \emph{{IEEE} J. Lightw. Technol.}, vol.~36, no.~20, pp.
  4843--4855, Oct 2018.

\bibitem{shahkarami2022complexity}
A.~Shahkarami, M.~Yousefi, and Y.~Jaouen, ``Complexity reduction over
  {Bi-RNN}-based nonlinearity mitigation in dual-pol fiber-optic communications
  via a {CRNN}-based approach,'' \emph{Opt.\ Fiber Technol.}, vol.~74, p.
  103072, Dec. 2022.

\bibitem{Freire:21}
P.~J. Freire, Y.~Osadchuk, B.~Spinnler, A.~Napoli, W.~Schairer, N.~Costa, J.~E.
  Prilepsky, and S.~K. Turitsyn, ``Performance versus complexity study of
  neural network equalizers in coherent optical systems,'' \emph{{IEEE} J.
  Lightw. Technol.}, vol.~39, no.~19, pp. 6085--6096, Oct 2021.

\bibitem{catanese2020fully}
C.~Catanese, R.~Ayassi, E.~Pincemin, and Y.~Jaou{\"e}n, ``A fully connected
  neural network approach to mitigate fiber nonlinear effects in 200g
  {DP-16-QAM} transmission system,'' in \emph{Int.\ Conf.\ Transparent Opt.\
  Netw.}, 2020, pp. 1--4.

\bibitem{Sidelnikov:21}
O.~Sidelnikov, A.~Redyuk, S.~Sygletos, M.~Fedoruk, and S.~Turitsyn, ``Advanced
  convolutional neural networks for nonlinearity mitigation in long-haul wdm
  transmission systems,'' \emph{{IEEE} J. Lightw. Technol.}, vol.~39, no.~8,
  pp. 2397--2406, Apr 2021.

\bibitem{darweesh2022ecoc}
J.~Darweesh, N.~Costa, A.~Napoli, B.~Spinnler, Y.~Jaou{\"e}n, and M.~Yousefi,
  ``Few-bit quantization of neural networks for nonlinearity mitigation in a
  fiber transmission experiment,'' in \emph{Eur.\ Conf.\ Opt.\ Commun.\ Conf.},
  Sep. 2022, pp. 1--4.

\bibitem{ignatova2019}
A.~Ignatov, R.~Timofte, A.~Kulik, S.~Yang, K.~Wang, F.~Baum, M.~Wu, L.~Xu, and
  L.~Van~Gool, ``Ai benchmark: {A}ll about deep learning on smartphones in
  2019,'' in \emph{IEEE/CVF Int.\ Conf.\ Comput.\ Vis.\ Workshops}, Oct. 2019,
  pp. 3617--3635.

\bibitem{Shortt:06}
A.~E. Shortt, T.~J. Naughton, and B.~Javidi, ``A companding approach for
  nonuniform quantization of digital holograms of three-dimensional objects,''
  \emph{Opt. Exp.}, vol.~14, no.~12, pp. 5129--5134, Jun 2006.

\bibitem{han2015learning}
S.~Han, J.~Pool, J.~Tran, and W.~Dally, ``Learning both weights and connections
  for efficient neural network,'' \emph{Adv.\ Neural Info.\ Process.\ Syst.},
  vol.~28, Dec. 2015.

\bibitem{lin2016pot}
Z.~Lin, M.~Courbariaux, R.~Memisevic, and Y.~Bengio, ``Neural networks with few
  multiplications,'' \emph{The Int.\ Conf.\ Learn. Rep.}, May 2016.

\bibitem{li2019}
Y.~Li, X.~Dong, and W.~Wang, ``Additive powers-of-two quantization: An
  efficient non-uniform discretization for neural networks,''
  \emph{arXiv:1909.13144}, 2019.

\bibitem{elhoushi2021deepshift}
M.~Elhoushi, Z.~Chen, F.~Shafiq, Y.~H. Tian, and J.~Y. Li, ``Deepshift:
  {}towards multiplication-less neural networks,'' in \emph{IEEE/CVF Conf.\
  Comput.\ Vision Pattern Recognit.}, Jun. 2021, pp. 2359--2368.

\bibitem{yamamoto2021learnable}
K.~Yamamoto, ``Learnable companding quantization for accurate low-bit neural
  networks,'' in \emph{IEEE/CVF Conf.\ Comput.\ Vision Pattern Recognit.}, Jun.
  2021, pp. 5029--5038.

\bibitem{peric2021robust}
Z.~Peric, B.~Denic, M.~Dincic, and J.~Nikolic, ``Robust 2-bit quantization of
  weights in neural network modeled by {Laplacian} distribution,'' \emph{Adv.
  Electr. Comput. Eng}, vol.~21, pp. 3--10, Aug. 2021.

\bibitem{Yamamoto_2021_CVPR}
K.~Yamamoto, ``Learnable companding quantization for accurate low-bit neural
  networks,'' in \emph{IEEE/CVF Conf.\ Comput.\ Vision Pattern Recognit.}, Jun.
  2021, pp. 5029--5038.

\bibitem{qu2020adaptive}
Z.~Qu, Z.~Zhou, Y.~Cheng, and L.~Thiele, ``Adaptive loss-aware quantization for
  multi-bit networks,'' in \emph{IEEE/CVF Conf.\ Comput.\ Vision Pattern
  Recognit.}, Jun. 2020, pp. 7988--7997.

\bibitem{NEURIPS2020_d77c7035}
Z.~Dong, Z.~Yao, D.~Arfeen, A.~Gholami, M.~W. Mahoney, and K.~Keutzer,
  ``{HAWQ-V2}: {Hessian} aware trace-weighted quantization of neural
  networks,'' in \emph{Adv.\ Neural Info.\ Process.\ Syst.}, vol.~33, Dec.
  2020, pp. 18\,518--18\,529.

\bibitem{choukroun}
Y.~Choukroun, E.~Kravchik, F.~Yang, and P.~Kisilev, ``Low-bit quantization of
  neural networks for efficient inference,'' in \emph{IEEE/CVF Int.\ Conf.\
  Comput.\ Vis.\ Workshops}, Dec. 2019, pp. 3009--3018.

\bibitem{hubara2020improving}
I.~Hubara, Y.~Nahshan, Y.~Hanani, R.~Banner, and D.~Soudry, ``Improving post
  training neural quantization: Layer-wise calibration and integer
  programming,'' \emph{arXiv:2006.10518v2}, Dec. 2020.

\bibitem{pmlr-v119-nagel20a}
M.~Nagel, R.~A. Amjad, M.~Van~Baalen, C.~Louizos, and T.~Blankevoort, ``Up or
  down? {A}daptive rounding for post-training quantization,'' in \emph{Int.\
  Conf.\ Mach.\ Learn.}, vol. 119.\hskip 1em plus 0.5em minus 0.4em\relax PMLR,
  13--18 Jul 2020, pp. 7197--7206.

\bibitem{NEURIPS2019_c0a62e13}
R.~Banner, Y.~Nahshan, and D.~Soudry, ``Post training 4-bit quantization of
  convolutional networks for rapid-deployment,'' in \emph{Adv.\ Neural Info.\
  Process.\ Syst.}, vol.~32, Dec. 2019.

\bibitem{liu2018bi}
Z.~Liu, B.~Wu, W.~Luo, X.~Yang, W.~Liu, and K.-T. Cheng, ``Bi-real net:
  Enhancing the performance of 1-bit cnns with improved representational
  capability and advanced training algorithm,'' in \emph{Euro Conf. Comp.
  Vision}, 2018, pp. 722--737.

\bibitem{zhou2016dorefa}
S.~Zhou, Y.~Wu, Z.~Ni, X.~Zhou, H.~Wen, and Y.~Zou, ``Dorefa-net: {T}raining
  low bitwidth convolutional neural networks with low bitwidth gradients,''
  \emph{arXiv:1606.06160}, 2016.

\bibitem{liu2019learning}
Z.-G. Liu and M.~Mattina, ``Learning low-precision neural networks without
  straight-through estimator ({STE}),'' \emph{arXiv:1903.01061}, 2019.

\bibitem{zhou2017incremental}
A.~Zhou, A.~Yao, Y.~Guo, L.~Xu, and Y.~Chen, ``Incremental network
  quantization: Towards lossless {CNNs} with low-precision weights,'' in
  \emph{The Int.\ Conf.\ Learning Rep.}, Apr. 2017, pp. 1--24.

\bibitem{darweesh2023oecc}
J.~Darweesh, N.~Costa, Y.~Jaou{\"e}n, A.~Napoli, B.~Spinnler, and M.~Yousefi,
  ``Successive quantization of the neural network equalizers in optical fiber
  communication,'' in \emph{OptoElectron.\ Commun.\ Conf.}, Jun. 2023, pp.
  1--6.

\bibitem{baskin2021uniq}
C.~Baskin, N.~Liss, E.~Schwartz, E.~Zheltonozhskii, R.~Giryes, A.~M. Bronstein,
  and A.~Mendelson, ``{UNIQ}: {U}niform noise injection for non-uniform
  quantization of neural networks,'' \emph{ACM Trans.\ Comput. Sys.}, vol.~37,
  no. 1-4, pp. 1--15, Mar. 2021.

\bibitem{liu2021fpga}
L.~Liu, X.~Liu, Z.~Zhai, Y.~Wu, H.~Jiang, L.~Yi, W.~Hu, and Q.~Zhuge,
  ``{FPGA}-based implementation of artificial neural network for nonlinear
  signal-to-noise ratio estimation,'' in \emph{OptoElectron.\ Commun.\ Conf.},
  Jul. 2021, pp. T2B--4.

\bibitem{kaneda2022}
N.~Kaneda, C.-Y. Chuang, Z.~Zhu, A.~Mahadevan, B.~Farah, K.~Bergman,
  D.~Van~Veen, and V.~Houtsma, ``Fixed-point analysis and {FPGA} implementation
  of deep neural network based equalizers for high-speed {PON},'' \emph{{IEEE}
  J. Lightw. Technol.}, vol.~40, no.~7, pp. 1972--1980, Apr. 2022.

\bibitem{fougstedt2018asic}
C.~Fougstedt, C.~H{\"a}ger, L.~Svensson, H.~D. Pfister, and P.~Larsson-Edefors,
  ``Asic implementation of time-domain digital backpropagation with
  deep-learned chromatic dispersion filters,'' in \emph{Eur.\ Conf.\ Opt.\
  Commun.\ Conf.}\hskip 1em plus 0.5em minus 0.4em\relax IEEE, Sep. 2018, pp.
  1--3.

\bibitem{he2021quant}
P.~He, F.~Wu, M.~Yang, A.~Yang, P.~Guo, Y.~Qiao, and X.~Xin, ``A fiber
  nonlinearity compensation scheme with complex-valued dimension-reduced neural
  network,'' \emph{IEEE Photon. J.}, vol.~13, no.~6, pp. 1--7, Oct. 2021.

\bibitem{freire2022hardware}
D.~A. Ron, P.~J. Freire, J.~E. Prilepsky, M.~Kamalian-Kopae, A.~Napoli, and
  S.~K. Turitsyn, ``Experimental implementation of a neural network optical
  channel equalizer in restricted hardware using pruning and quantization,''
  \emph{Scientific Reports}, vol.~12, no.~1, p. 8713, May 2022.

\bibitem{fougstedt2018quant}
C.~Fougstedt, L.~Svensson, M.~Mazur, M.~Karlsson, and P.~Larsson-Edefors,
  ``{ASIC} implementation of time-domain digital back propagation for coherent
  receivers,'' \emph{{IEEE} Photon. Technol. Lett.}, vol.~30, no.~13, pp.
  1179--1182, Jul. 2018.

\bibitem{freire2023reducing}
P.~J. Freire, A.~Napoli, D.~A. Ron, B.~Spinnler, M.~Anderson, W.~Schairer,
  T.~Bex, N.~Costa, S.~K. Turitsyn, and J.~E. Prilepsky, ``Reducing
  computational complexity of neural networks in optical channel equalization:
  From concepts to implementation,'' \emph{{IEEE} J. Lightw. Technol.}, Jan.
  2023.

\bibitem{huang2022low}
X.~Huang, D.~Zhang, X.~Hu, C.~Ye, and K.~Zhang, ``Low-complexity recurrent
  neural network based equalizer with embedded parallelization for
  100-{Gbit/s}/$\lambda$ {PON},'' \emph{{IEEE} J. Lightw. Technol.}, vol.~40,
  no.~5, pp. 1353--1359, Mar. 2022.

\bibitem{chagnon2019quant}
M.~Chagnon, J.~Siirtola, T.~Rissa, and A.~Verma, ``Quantized deep neural
  network empowering an {IM-DD} link running in realtime on a field
  programmable gate array,'' in \emph{Opt.\ Fiber Conf.}, Mar. 2019.

\bibitem{xu2019efficient}
W.~Xu, X.~Tan, Y.~Lin, X.~You, C.~Zhang, and Y.~Be’ery, ``On the efficient
  design of neural networks in communication systems,'' in \emph{Asilomar
  Conf.\ Signals, Syst., and Comput.}\hskip 1em plus 0.5em minus 0.4em\relax
  IEEE, Nov. 2019, pp. 522--526.

\bibitem{koike2021}
T.~Koike-Akino, Y.~Wang, K.~Kojima, K.~Parsons, and T.~Yoshida,
  ``Zero-multiplier sparse {DNN} equalization for fiber-optic qam systems with
  probabilistic amplitude shaping,'' in \emph{Eur.\ Conf.\ Opt.\ Commun.\
  Conf.}, 2021, pp. 1--4.

\bibitem{aoudia2019quant}
F.~A. Aoudia and J.~Hoydis, ``Towards hardware implementation of neural
  network-based communication algorithms,'' in \emph{IEEE Int. Workshop Signal
  Process.\ Adv.\ Wireless Commun.}, Jul. 2019, pp. 1--5.

\bibitem{zhuge2014quant}
Q.~Zhuge, M.~Reimer, A.~Borowiec, M.~O'Sullivan, and D.~V. Plant, ``Aggressive
  quantization on perturbation coefficients for nonlinear pre-distortion,'' in
  \emph{Opt.\ Fiber Conf.}, Mar. 2014, pp. 1--3.

\bibitem{darabi2020regularized}
S.~Darabi, M.~Belbahri, M.~Courbariaux, and V.~P. Nia, ``Regularized binary
  network training,'' \emph{arXiv:1812.11800v3}, Apr. 2020.

\bibitem{liu2022ternary}
B.~Liu, F.~Li, X.~Wang, B.~Zhang, and J.~Yan, ``Ternary weight networks,'' in
  \emph{IEEE Int.\ Conf.\ Acoustics, Speech and Signal Proc.}, Jun. 2023, pp.
  1--5.

\bibitem{zhou2018dorefanet}
S.~Zhou, Y.~Wu, Z.~Ni, X.~Zhou, H.~Wen, and Y.~Zou, ``{DoReFa-Net}: Training
  low bitwidth convolutional neural networks with low bitwidth gradients,''
  \emph{arXiv:1606.06160v3}, Feb. 2018.

\bibitem{freire2022complexity}
P.~J. Freire, S.~Srivallapanondh, A.~Napoli, J.~E. Prilepsky, and S.~K.
  Turitsyn, ``Computational complexity evaluation of neural network
  applications in signal processing,'' \emph{arXiv:2206.12191}, pp. 1--13,
  2022.

\end{thebibliography}

%%%%%%%%%%%%%%%%%%%%%%%%% Refs

\end{document}